\documentclass[lettersize,journal]{IEEEtran}
\usepackage{amsmath,amsfonts}
\usepackage{array}
\usepackage{subcaption}
\usepackage{textcomp}
\usepackage{stfloats}
\usepackage{url}
\usepackage{verbatim}
\usepackage{graphicx}
\usepackage{cite}
\usepackage{color}
\usepackage{makecell}
\usepackage{pifont}

\usepackage{amsthm}
\usepackage[ruled,vlined,linesnumbered]{algorithm2e}
\usepackage{wrapfig}
\usepackage{graphicx}

\newcommand{\cmark}{\ding{51}}%
\newcommand{\xmark}{\ding{55}}%

\usepackage{adjustbox}      
\usepackage{booktabs}

\usepackage{amsmath}
\DeclareMathOperator*{\argmin}{arg\,min}

\begin{document}

\title{IAPRepair: In-Network Aggregation Enhanced Proactive Repair for Erasure-Coded Storage System}

\author{Lei Liu,~Yong Wang, ~Junqi Chen,~\IEEEmembership{Member, IEEE}, ~Yangfan Liang,~\IEEEmembership{Member, IEEE}, ~Qian He, ~Baokang Zhao


\thanks{
This work was supported by the “Jianfeng” Action Plan of the Guangxi Science and Technology Major Program (Grant No. 2026JF0601001-2), the Guangxi Key Research and Development Program, China (Grant No. AB25069178), the Open Project Program of Guangxi Key Laboratory of Digital Infrastructure (Grant No. GXDIOP2024008), the Project for Enhancing Young and Middleaged Teacher’s Research Basis Ability in Colleges of Guangxi (Grant No. 2025KY0250), the National Natural Science Foundation of China (Grant Nos. 62461057, 61861013), and the Innovation Project of GUET Graduate Education (Grant Nos. 2025YCXS059, 2026YCXS062) (\textit{Corresponding author: Junqi Chen.})
}
\thanks{
Lei Liu is with the School of Computer Science and Information Security, Guilin University of Electronic Technology, Guilin 541004, China, and also with the School of Computer Science and Engineering, Guilin University of Aerospace Technology, Guilin 541004, China (e-mail: leiz2018@guat.edu.cn).
}
\thanks{
Yong Wang, Junqi Chen are with the School of Computer Science and Information Security, Guilin University of Electronic Technology, Guilin 541004, China  (e-mail: ywang@guet.edu.cn; cookiechen@mails.guet.edu.cn).
}
\thanks{
Yangfan Liang is with the College of Information Science and Engineering, Jiaxing University, Jiaxing 314001, China
}
 \thanks{
 Qian He  are with the Guangxi University Engineering Research Center of Cloud Network Convergence and Data Security, Guilin University of Electronic Technology, Guilin 541004, China (e-mail: heqian@guet.edu.cn; cookiechen@mails.guet.edu.cn).
}
\thanks{
B. Zhao is with the College of Computer Science and Technology, National University of Defense Technology, Changsha 410073, China.
}

}

\maketitle

\begin{abstract}
Erasure-coded storage provides fault tolerance with substantially lower storage overhead than full replication, but repairing lost or at-risk blocks requires intensive cross-node data transfer. Existing reactive repair schemes start only after a failure, while proactive schemes can move data before failure but commonly treat migration, reconstruction, and network aggregation as loosely coupled operations. As a result, receiver-side bottlenecks, heterogeneous available bandwidth, and limited programmable-switch state continue to constrain repair parallelism. This paper presents IAPRepair, an in-network aggregation enhanced proactive repair framework for erasure-coded storage. IAPRepair jointly constructs each repair batch, assigns reconstruction providers and replacement nodes according to normalized transmission loads, schedules migration around the remaining receive capacity, and selectively enables in-network aggregation for reconstruction blocks that would otherwise overload healthy nodes. The selective design reduces receiver-side traffic while retaining migration parallelism and respecting a configurable switch-resource budget. We implement IAPRepair with a Tofino programmable switch and 16 storage nodes, and evaluate it using both a prototype testbed and large-scale simulations. Across coding parameters, block sizes, node populations, and multiple STF-node scenarios, IAPRepair reduces repair time by at least 47.83\% compared with the evaluated state-of-the-art methods. The results demonstrate that coordinating proactive repair decisions with in-network processing is an effective way to improve repair efficiency under bandwidth heterogeneity.
\end{abstract}


\section{Introduction}
\label{sec:introduction}

Large-scale distributed storage systems increasingly rely on erasure coding to provide fault tolerance with much lower storage overhead than full replication. Reed--Solomon (RS) codes, for example, encode $k$ data blocks into $n$ blocks and allow a missing block to be reconstructed from any $k$ surviving blocks. The benefit comes with a substantial repair cost: reconstructing one block requires data to be retrieved from multiple nodes, and a node failure creates many concurrent block-repair tasks. The resulting cross-node traffic prolongs the data-unavailability window and can directly affect storage-system reliability.

Most existing repair systems are reactive. Pipeline-based methods such as PR, PPT, and PivotRepair reduce incast effects by coordinating transmissions across storage nodes, while RepairBoost and SelectiveEC balance repair workloads after a failure has occurred \cite{li2021repair,bai2019fast,yao2022pivotrepair,lin2021boosting,xu2021selectiveec}. Failure prediction makes it possible to repair data before a node becomes unavailable. We refer to a node predicted to fail as a soon-to-fail (STF) node. Proactive repair can either migrate STF blocks directly to healthy nodes or reconstruct them from surviving blocks, and recent systems such as FastPR and LFPR combine these two operations in batches \cite{zhu2013proactive,botezatu2016predicting,mahdisoltani2017proactive,zhang2020hddse,sit2006proactive,li2022fast,wu2022lfpr}. Nevertheless, current proactive schemes do not sufficiently coordinate receiver capacity, bandwidth heterogeneity, and the choice between migration and reconstruction.

The central difficulty is an asymmetric traffic pattern. Both STF and healthy nodes may provide repair data, but only healthy nodes can receive repaired blocks. Concurrent migration and reconstruction can therefore overload a small subset of healthy receivers. The problem is amplified when the available bandwidth of storage nodes is shared with front-end workloads and differs across nodes. A repair schedule that is balanced in terms of block counts can still be highly imbalanced in transmission time.

\begin{figure*}[t]
\centering
\includegraphics[width=0.99\textwidth]{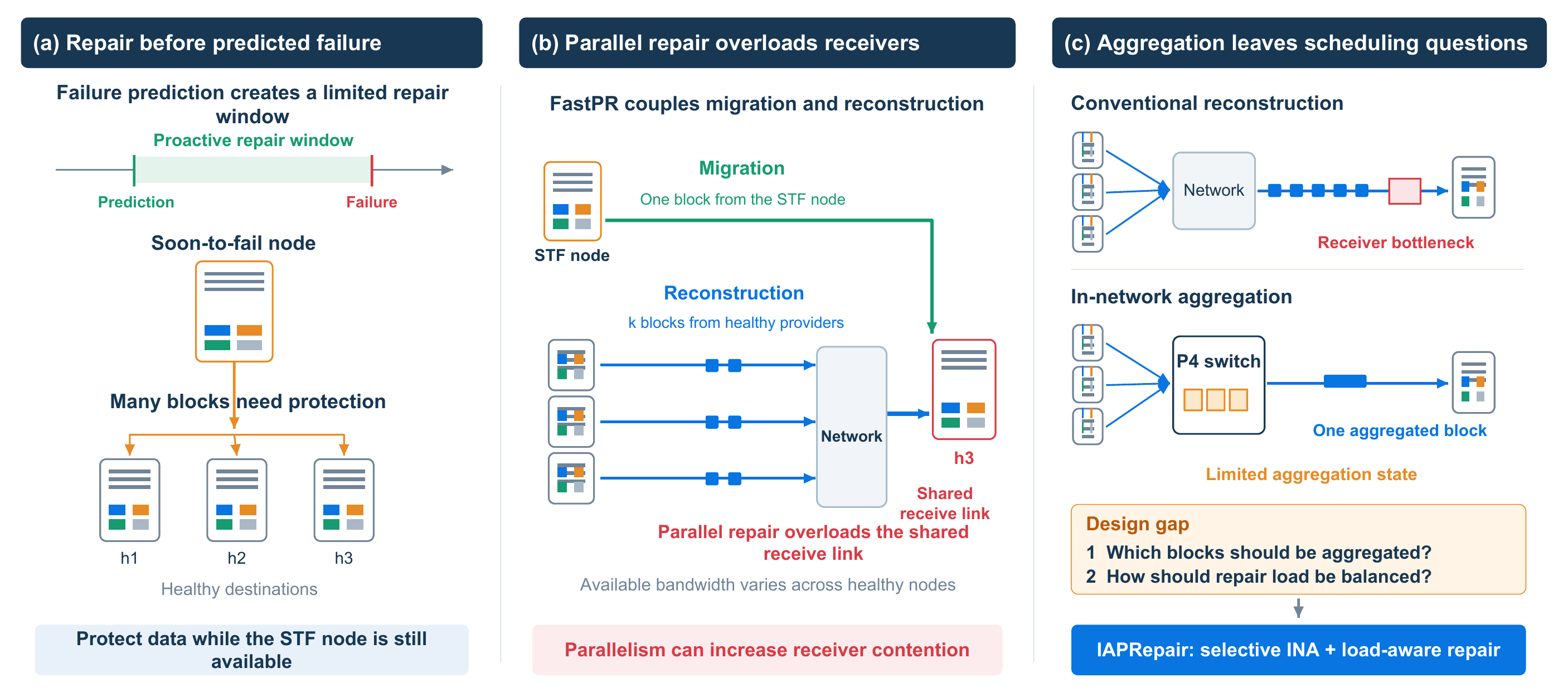}
\caption{Motivation for IAPRepair. Predicted failure creates a repair burst; concurrent migration and reconstruction can overload healthy receivers; and naive in-network aggregation alone does not determine which blocks to aggregate or how to balance the remaining repair load.}
\label{fig:motivation}
\end{figure*}

Programmable switches provide a complementary opportunity. In-network aggregation (INA) can combine reconstruction traffic on the forwarding path and reduce the volume delivered to storage nodes. Existing INA-based repair systems, including NetEC, Paint, and InaPR, primarily target reactive repair \cite{qiao2022netec,xia2024parallelized,liu2025network}. Applying INA to proactive repair, however, raises two coupled questions: which reconstruction blocks should consume scarce switch state, and how should migration and reconstruction tasks be assigned when node bandwidth is heterogeneous? Figure~\ref{fig:motivation} summarizes this design gap.

To address these issues, we present IAPRepair, an in-network aggregation enhanced proactive repair framework for erasure-coded storage. IAPRepair coordinates three decisions within each repair batch. First, it selects reconstruction blocks and assigns their provider and replacement nodes according to normalized uplink and downlink loads. Second, it schedules migration blocks around the residual receive capacity of healthy nodes. Third, it selectively enables INA for reconstruction blocks whose traffic would otherwise create a receiver bottleneck, while respecting a configurable switch-resource budget. This design couples network processing with load-aware repair scheduling, allowing migration and reconstruction to progress in parallel without indiscriminately consuming switch state.

The contributions of this paper are as follows.

\begin{itemize}
    \item We present IAPRepair, a proactive repair framework that combines migration, reconstruction, and selective in-network aggregation for erasure-coded storage.
    \item We develop a load-aware batch-construction strategy that jointly assigns reconstruction and migration tasks under heterogeneous node bandwidth and receiver-capacity constraints.
    \item We design a selective INA decision mechanism that reduces receiver-side traffic while limiting the number of reconstruction blocks processed by the programmable switch.
    \item We implement IAPRepair with a Tofino programmable switch and 16 storage nodes. Prototype and simulation results show that IAPRepair reduces repair time by at least 47.83\% compared with the evaluated state-of-the-art methods.
\end{itemize}

The rest of this paper is organized as follows. Section~\ref{sec:background} reviews erasure-coded repair, proactive repair, programmable-switch aggregation, and related work. Section~\ref{sec:architecture} presents the system architecture and problem formulation. Section~\ref{sec:Design} details the batch-construction and selective-aggregation algorithms. Section~\ref{sec:implementation} describes the prototype implementation. Section~\ref{sec:evaluation} reports the evaluation results, and Section~\ref{sec:Conclusion} concludes the paper.

\section{Background and Related Work}
\label{sec:background}
\subsection{Basics of erasure coding}

\begin{figure}[hb!]
\centering
\includegraphics[width=0.83\columnwidth]{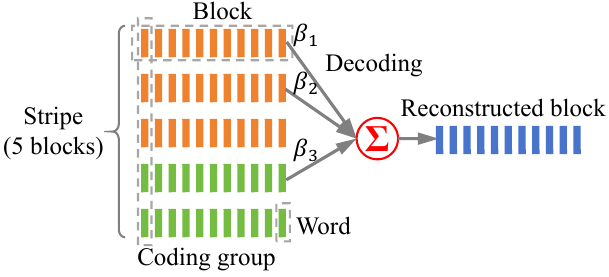}
\caption{Illustration of reconstruction with $RS(5,3)$}
\label{fig_reconstruction}
\end{figure}

Many erasure codes are proposed in the literature. In this paper, we focus on Reed–Solomon codes (denoted as $RS(n,k)$), which are widely deployed in production storage systems \cite{shen2025survey}. $RS(n,k)$ performs linear transformations on $k$ data blocks to generate $n-k$ parity blocks, and combines these $n$ blocks to form a stripe. A block can be reconstructed from any other $k$ blocks within the same stripe. Specifically, denoting  the failed block as $D^{*}$ and the selected blocks used to reconstruct $D^{*}$ as $D_{1},D_{2},\dots,D_{k}$, the reconstruction process can be represented as follows:
\begin{equation}
D^*=\sum_{i=1}^k \beta_{i} D_{i}
\label{eq_decode}
\end{equation}
\noindent where $\beta_i$ is the decoding coefficient of block $D_i$. The linear transformation is performed based on the Galois Field $GF(2^{w})$ \cite{greenan2008optimizing} over $w$-bit words. Fig. \ref{fig_reconstruction} illustrates data reconstruction using $RS(5,3)$ as an example. The reconstructed block can be any one of the five blocks.

Each block consists of a large number of $w$-bit words. Within a stripe, the words at the same offset across all blocks form an coding group. Decoding of different coding groups are independent of each other. Therefore, during the reconstruction of a failed block, a word can be reconstructed as long as any other $k$ words from the same coding group are available. This is also the theoretical basis for performing erasure-coded data repair using programmable switches, since buffering a whole block prior to data repair would exhaust the switch’s limited SRAM resources \cite{cheng2025sina}.

\subsection{Proactive repair}
\begin{figure}[ht!]
\centering
\includegraphics[width=0.9\columnwidth]{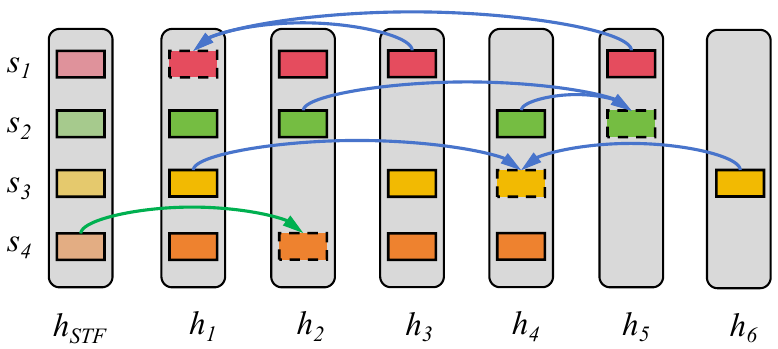}
\caption{Proactive repair via migration and reconstruction}
\label{fig_Proactive}
\end{figure}

Proactive repair, based on failure prediction, repairs the blocks stored on STF nodes. Compared to reactive repair, proactive repair exhibits two advantages. (1) Conducting data repair prior to failure occurrence achieves consistent data availability and enhances the reliability of the storage system. (2) Repair can be performed simultaneously via both migration and reconstruction, achieving higher repair efficiency. As illustrated in Fig. \ref{fig_Proactive}, the number of healthy nodes is typically much larger than that of STF nodes. Migration (the green line) directly copies a block from the STF node to the replacement node, which incurs a low data transmission overhead. However, since only the STF node can supply data for migration, the repair parallelism is inherently limited. Reconstruction (the blue lines) retrieves blocks from multiple healthy nodes. Although reconstruction enables parallel repair of multiple blocks, it requires a larger volume of data to reconstruct a single block. Coordinating migration and reconstruction through effective repair task scheduling can improve repair efficiency.

\subsection{In-network aggregation}
In-network aggregation is a technique that offloads data aggregation tasks (e.g., summation, averaging, counting, etc.) from severs to network devices. Aggregating the repair traffic directly on network devices along the transfer paths can reduce network load and alleviate bandwidth bottlenecks. Unlike traditional fixed-function switches, the Intel Tofino programmable switch \cite{agrawal2020intel} enables users to execute custom computation, storage and logical processing operations during packet forwarding via the P4 programming language \cite{bosshart2014p4}. The arithmetic logic units and SRAM-based stateful objects (e.g., registers) within the programmable switch can be utilized to implement data aggregation. Since erasure-coded data repair requires aggregating data from multiple nodes, the programmable switch can be leveraged to enhance the data reconstruction efficiency. However, in-network aggregation with programmable switches faces challenges such as scarcity of switch internal resources and the impact of uncertainty in packet arrival time on aggregation performance \cite{qiao2022netec, xia2024parallelized}.

\subsection{Related work}

Existing repair techniques can be organized along two dimensions: whether repair is triggered after a failure or before a predicted failure, and whether computation remains on storage nodes or is partially offloaded to network hardware. Table~\ref{tab:related_work} summarizes the distinctions most relevant to IAPRepair.

Reactive repair methods operate after a failure has been observed. PR partitions repair traffic into a pipeline, whereas PPT and PivotRepair refine the pipeline structure to reduce incast and improve transmission overlap \cite{li2021repair,bai2019fast,yao2022pivotrepair}. RepairBoost, SelectiveEC, SMFRepair, and SARepair further address workload placement or repair-stripe selection to balance node-side traffic \cite{lin2021boosting,xu2021selectiveec,zhou2022bandwidth,zhou2024stripe}. These methods improve reactive repair efficiency, but they do not exploit the additional time available before a predicted failure.

Proactive repair methods use failure prediction to move data before a node becomes unavailable. Proactive Replication and ProCode reduce the risk of a future failure through migration or replication, at the cost of additional traffic or storage overhead \cite{duminuco2007proactive,li2016procode}. FastPR couples migration with reconstruction in a batched workflow, and LFPR extends the idea with a lazy strategy for mobile distributed storage \cite{li2022fast,wu2022lfpr}. These approaches establish the basis for IAPRepair, but their scheduling decisions do not jointly account for receiver-side contention, heterogeneous available bandwidth, and limited in-network state.

Programmable-network repair methods move part of the repair computation into the network. NetEC and INC-EC offload decoding or decoding-related operations to programmable switches, while AggreTree, XORInc, Paint, and InaPR use host--switch collaboration to aggregate repair traffic \cite{qiao2022netec,jiang2023accelerating,xia2020efficient,wang2019xorinc,xia2024parallelized,liu2025network}. SmartNIC- and accelerator-based designs provide another hardware-assisted direction \cite{Shi9355252,MAISTO2026108187}. These methods reduce data movement, but they primarily target reactive repair and do not solve the joint scheduling problem of proactive migration and reconstruction. IAPRepair fills this gap by coupling load-aware proactive batch construction with selective INA under an explicit switch budget.

\begin{table*}[t]
\centering
\caption{Comparison of repair approaches for erasure-coded storage. INA denotes in-network aggregation.}
\label{tab:related_work}
\resizebox{\textwidth}{!}{%
\begin{tabular}{p{2.7cm}p{1.35cm}p{4.0cm}p{1.0cm}p{4.1cm}p{1.2cm}}
\toprule
\textbf{Representative work} & \textbf{Trigger} & \textbf{Core repair strategy} & \textbf{INA} & \textbf{Bandwidth/load awareness} & \textbf{Proactive} \\
\midrule
PR, PPT, PivotRepair \cite{li2021repair,bai2019fast,yao2022pivotrepair} & Reactive & Pipelined repair and transmission overlap & \xmark & Pipeline-oriented balancing; no proactive receiver model & \xmark \\
RepairBoost, SelectiveEC, SMFRepair, SARepair \cite{lin2021boosting,xu2021selectiveec,zhou2022bandwidth,zhou2024stripe} & Reactive & Load balancing, forwarding, or stripe-aware repair & \xmark & Node-side workload balancing & \xmark \\
Proactive Replication, ProCode \cite{duminuco2007proactive,li2016procode} & Proactive & Migration or replication before failure & \xmark & Fixed-rate or replication-oriented scheduling & \cmark \\
FastPR, LFPR \cite{li2022fast,wu2022lfpr} & Proactive & Batched migration combined with reconstruction & \xmark & Limited joint handling of heterogeneous load & \cmark \\
NetEC, INC-EC \cite{qiao2022netec,jiang2023accelerating} & Reactive & Switch-assisted decoding or reconstruction & \cmark & Constrained by switch computation and state & \xmark \\
AggreTree, XORInc, Paint, InaPR \cite{xia2020efficient,wang2019xorinc,xia2024parallelized,liu2025network} & Reactive & Host--switch collaborative aggregation & \cmark & Sensitive to switch state and packet-arrival skew & \xmark \\
\textbf{IAPRepair} & \textbf{Proactive} & \textbf{Joint batch selection for migration and reconstruction with selective INA} & \textbf{\cmark} & \textbf{Uplink/downlink load balancing under a switch budget} & \textbf{\cmark} \\
\bottomrule
\end{tabular}}
\end{table*}

\section{SYSTEM ARCHITECTURE AND PROBLEM FORMULATION}
\label{sec:architecture}\label{sec:Formulation}
\subsection{System overview}

\begin{figure*}[t]
\centering
\includegraphics[width=0.99\textwidth]{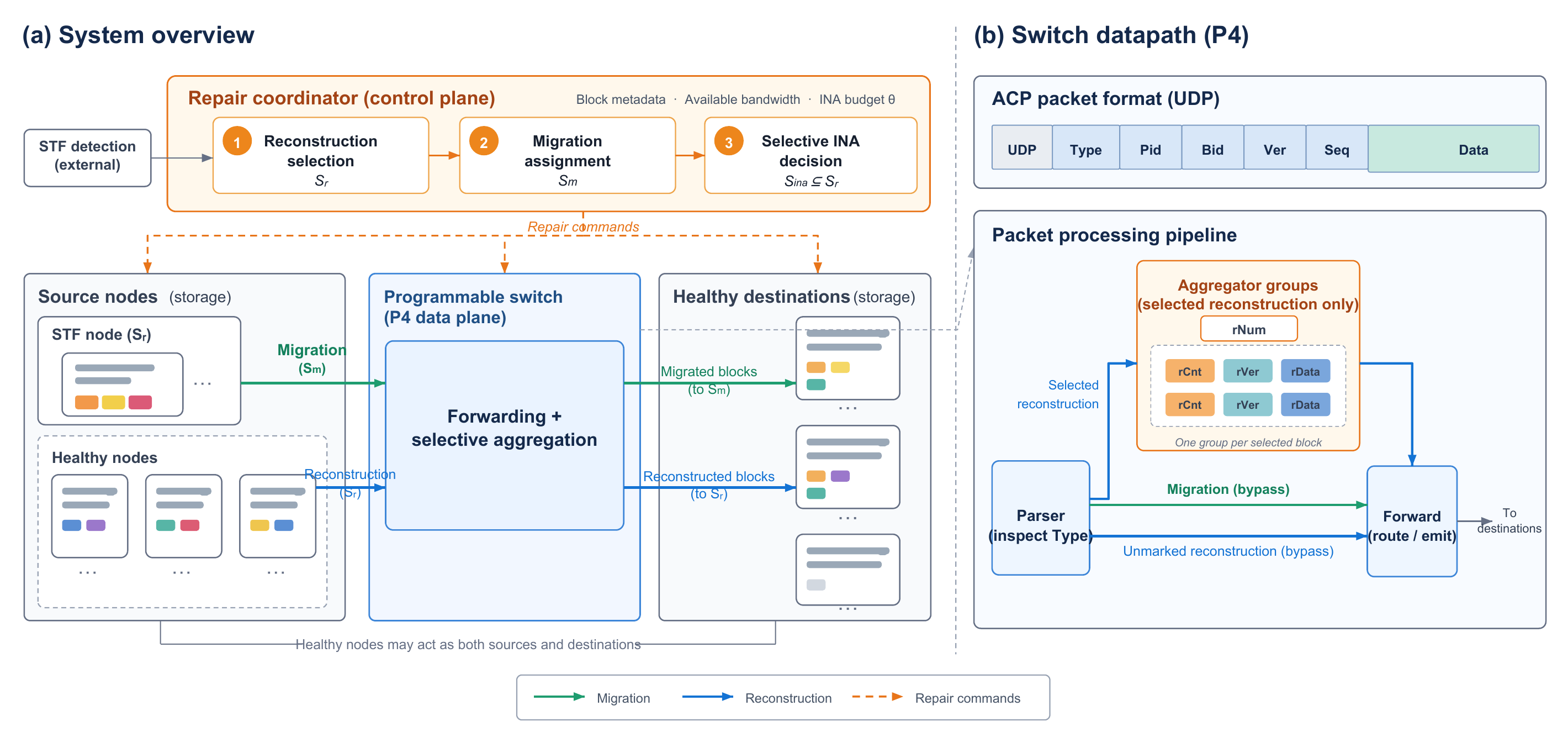}
\caption{Overall architecture of IAPRepair. The control plane coordinates reconstruction selection, migration assignment, and selective INA; the P4 data plane parses ACP packets, forwards unmarked traffic, and aggregates selected reconstruction packets.}
\label{fig:overall_architecture}
\end{figure*}

IAPRepair uses a control plane to select repair blocks and assign providers and destinations, and a programmable-switch data plane to forward packets and aggregate only the selected reconstruction traffic. The overall workflow is shown in Fig.~\ref{fig:overall_architecture}. The coordinator first determines the reconstruction set, then assigns migration blocks around the resulting receive load, and finally marks the subset of reconstruction blocks that should consume aggregator state. The data plane parses the packet type, forwards migration and unmarked reconstruction packets directly, and routes marked reconstruction packets through a dedicated aggregator group before emission to healthy destinations.

In this section, we describe and analyze the problems in proactive repair, and then formulate a model for proactive repair. Our work is built on existing failure prediction mechanisms \cite{zhu2013proactive, botezatu2016predicting, mahdisoltani2017proactive, zhang2020hddse}, and focuses on the proactive repair strategies after STF node detection.

\subsection{Problem description and Analysis}
Given that a large number of blocks are stored on a node, the proactive repair of an STF node is typically performed in batches to avoid excessive consumption of network and memory resources \cite{li2022fast,xu2021selectiveec}. Let $n_{m}$ and $n_{r}$ denote the number of blocks repaired via migration and reconstruction in a batch, respectively. Let $h_{STF}$ denote the STF node, $h_{i}$ denote the $i$-th healthy node, and $s_{j}$ denotes the $j$-th STF block. STF nodes can migrate their blocks to any healthy node but cannot receive data from other nodes. All healthy nodes can serve as replacement nodes and receive migration and reconstruction data from other nodes.


Problem 1: How to alleviate the downlink bottlenecks in proactive repair?

\begin{figure*}[t]
\centering
\begin{subfigure}{.33\textwidth}
  \centering
  \includegraphics[width=0.95\columnwidth]{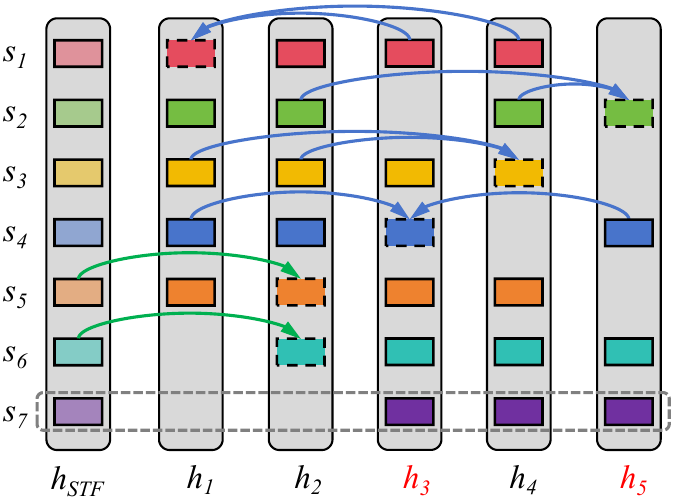}
  \caption{Without in-network aggregation}
\label{fig_problem_a}
\end{subfigure}%
\begin{subfigure}{.33\textwidth}
  \centering
  \includegraphics[width=0.95\columnwidth]{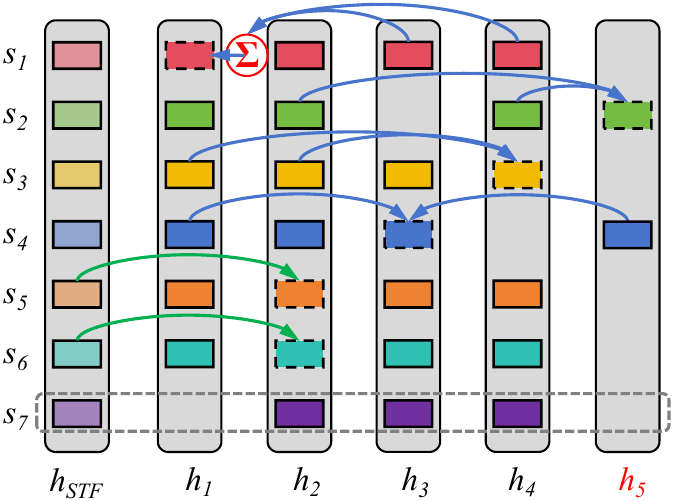}
  \caption{Skewed repair workload}
  \label{fig_problem_b}
\end{subfigure}%
\begin{subfigure}{.33\textwidth}
  \centering
  \includegraphics[width=0.95\columnwidth]{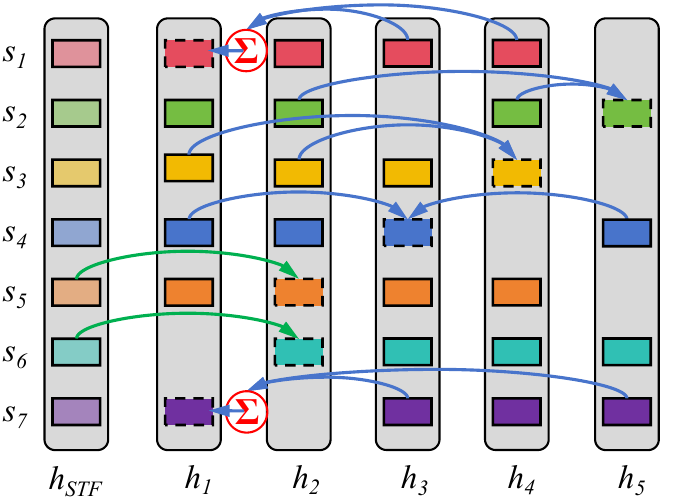}
  \caption{Repair with maximum efficiency}
  \label{fig_problem_c}
\end{subfigure}%
\caption{Comparison of proactive repair efficiency across different strategies. In (a) and (b), the number of STF nodes repaired is limited to 6. In (c), 7 STF stripes are repaired in parallel.}
\label{fig:sample_optimization}
\end{figure*}

All nodes can provide data to repair STF blocks. Specifically, STF nodes provide data for the migration of STF blocks, whereas healthy nodes provide data for the reconstruction of STF blocks. However, the repaired blocks can only be stored on healthy nodes. Consequently, the number of receiving nodes is smaller than that of sending nodes, which may overload the receiving nodes and degrade repair efficiency. As illustrated in Fig. \ref{fig_problem_a}, when the numbers of migrated and reconstructed blocks are 2 and 4, respectively, every healthy node needs to receive two data blocks, whereas nodes $h_{3}$ and $h_{5}$ send only one data block each. The receiving capacity limits repair performance.

In-network aggregation provides an solution to this problem. By reducing bandwidth consumption, in-network aggregation enables more STF blocks to be reconstructed within a single batch. As illustrated in Fig. \ref{fig_problem_c}, employing in-network aggregation to reconstruct $s_{1}$ and $s_{7}$ alleviates the bandwidth demand on $h_{1}$, enabling it to complete the reconstruction of two STF blocks. All nodes send and receive two blocks.

Problem 2: How to select an appropriate set of STF blocks for each batch?

The selection of STF blocks for a batch impacts the repair efficiency. Consider a batch comprising 7 STF blocks as illustrated in Fig. \ref{fig_problem_b}, where $s_{6}$ and $s_{7}$ differ from those in Fig. \ref{fig_problem_c}. Although in-network aggregation is employed for data reconstruction in Fig. \ref{fig_problem_b}, the repair efficiency is lower than that in Fig. \ref{fig_problem_c}. This is because $h_{5}$ can provide data for the reconstruction of only one STF block. Consequently, to repair $s_{7}$, at least one of $h_{STF}$, $h_{2}$, and $h_{4}$ must transmit three blocks. This uneven distribution of the transmission workload adversely affects repair performance. Furthermore, heterogeneity in available bandwidth across nodes exacerbates both the severity and complexity of this issue.

\subsection{Formulation}

Let $n_{STF}$ denote the total number of STF blocks, indexed by $S=\{1,2,\ldots,n_{STF}\}$. Let $S_{m}$ and $S_{r}$ denote the sets of STF blocks to be repaired via migration and reconstruction, respectively, in a batch. Let $H_{h}$ denote the set of healthy nodes, and $H_{STF}$ denote the set of STF nodes. The set of healthy nodes capable of providing data for the reconstruction of $s_{j}$ is denoted as $pro(s_{j})$. The decision variables $\varphi_{i,j}$ and $\phi_{i,j}$ are binary (Eq. \ref{eq_fml_varphi_val} and \ref{eq_fml_phi_val}). If $h{_i}$ provides a block for the reconstruction of $s_{j}$, $\varphi_{i,j}=1$, and $\varphi_{i,j}=0$ otherwise. Eq. \ref{eq_fml_varphi_eq_0_Sr} restricts that only healthy nodes in $pro(s_{j})$ are permitted to provide data for the reconstruction of STF block $s_{j}$.  Eq. \ref{eq_fml_varphi_eq_0_Sm} ensures that only STF nodes can provide data for migration. If $h_{i}$ is selected as the replacement node for $s_{j}$ to store the repaired STF block, $\phi_{i,j}=1$, and $\phi_{i,j}=0$ otherwise. Eqs. \ref{eq_fml_phi_eq_0} and \ref{eq_fml_phi_leq_1} require all repaired blocks to be stored exclusively on healthy nodes, and blocks belonging to the same stripe to be placed on distinct nodes for fault tolerance, where, $g(s_{j})$  denotes the stripe to which $s_{j}$  belongs.

\begin{equation}
    \varphi_{i,j}=\{0,1\},\forall i\in H_{h}\cap H_{STF},j\in S
\label{eq_fml_varphi_val}
\end{equation}

\begin{equation}
    \phi_{i,j}=\{0,1\}, \forall i\in H_{h},j\in S
\label{eq_fml_phi_val}
\end{equation}

\begin{equation}
    \varphi_{i,j}=0,\forall i\in(H_{h}\setminus pro(s_j))\cup H_{STF},j\in S_{r}
\label{eq_fml_varphi_eq_0_Sr}
\end{equation}

\begin{equation}
    \varphi_{i,j}=0,\forall i\in H_{h},j\in S_{m}
\label{eq_fml_varphi_eq_0_Sm}
\end{equation}

\begin{equation}
    \phi_{i,j}=0,\forall i\in pro(s_{j})\cup H_{STF}, j\in S_{r} \cup S_{m}
\label{eq_fml_phi_eq_0}
\end{equation}

\begin{equation}
    \phi_{i,j}+\phi_{i,j'}\leq 1,\forall g(s_{j})=g(s_{j'}),j\neq j'
\label{eq_fml_phi_leq_1}
\end{equation}

Reconstructing an STF block requires $k$ healthy nodes to provide data, whereas migrating an STF block requires only one STF node to provide data (Eqs. \ref{eq_fml_recon_k} and 
\ref{eq_fml_mgrt_1}).  Eq \ref{eq_fml_tgt} guarantees that each STF block is assigned a designated replacement node. 

\begin{equation}
    \sum_{i\in H_{h}}\varphi_{i,j}=k,\forall j\in S_{r}
\label{eq_fml_recon_k}
\end{equation}

\begin{equation}
    \sum_{i\in H_{STF}}\varphi_{i,j}=1,\forall j\in S_{m}
\label{eq_fml_mgrt_1}
\end{equation}

\begin{equation}
    \sum_{i\in H_{STF}}\phi_{i,j}=1,\forall j\in S_{r} \cup S_{m}
\label{eq_fml_tgt}
\end{equation}

The number of blocks transmitted by each node is denoted as $n_{i}^{u}$, computed via Eq. \ref{eq_fml_ntx}. Let $\psi_{j}$ be a binary indicator, where $\psi_{j}=1$ signifies that in-network aggregation is employed for the reconstruction of block $s_{j}$, and $\psi_{j}=0$ indicates that data aggregation is performed at the storage node (Eq. \ref{eq_fml_psi}). Without in-network aggregation, a node must receive $k$ blocks to reconstruct one block. In contrast, for the migration or in-network aggregation-enabled reconstruction, the node needs to receive only one block. The number of blocks received by each node is denoted as $n_{i}^{d}$, computed via Eq. \ref{eq_fml_nrx}. Due to limited switch resources, Eq. \ref{eq_fml_psi_sum} limits the number of blocks reconstructed through in-network aggregation, where $\theta$ denotes the maximum number of such blocks.

\begin{equation}
    n_{i}^{u}=\sum_{j\in S_{r}\cup S_{m}} \varphi_{i,j},\forall i\in H_{STF}\cup H_{h}
\label{eq_fml_ntx}
\end{equation}

\begin{equation}
    \psi_{j}\in\{0,1\},\forall j\in S_{r}
\label{eq_fml_psi}
\end{equation}

\begin{equation}
    n_{i}^{d}=\sum_{j\in S_{r}}\phi_{i,j}(k(1-\psi_{j})+\psi_{j})+\sum_{j\in S_{m}}\phi_{i,j},\forall i\in H_{h}
\label{eq_fml_nrx}
\end{equation}

\begin{equation}
    \sum_{j\in S_{r}}\psi_{j}\leq\theta
\label{eq_fml_psi_sum}
\end{equation}

Let $b_{size}n_{i}^{u}$ and $b_{size}n_{i}^{d}$ denote the data volumes transmitted and received by a node, respectively, where $b_{size}$ is the block size. Let $u_{i}$ and $d_{i}$ denote the available uplink and downlink bandwidths of $h_i$, respectively. The repair time of the entire batch, denoted by $t_{bch}$, is determined by the node with the maximum transmission time. $t_{bch}$ can be calculated using the Eq \ref{eq_fml_t_batch}. 

\begin{equation}
    \begin{aligned}
    t_{bch} = b_{size} \max\biggl( 
    &\left\{\frac{n_{i}^{u}}{u_{i}} \,\middle|\, i\in H_{h}\cup H_{STF}\right\} \cup {} \\
    &\left\{\frac{n_{i}^{d}}{d_i} \,\middle|\, i\in H_{h}\right\} \biggr)
\end{aligned}
\label{eq_fml_t_batch}
\end{equation}

IAPRepair aims to reduce repair time by maximizing the repair efficiency of each batch. The objective of IAPRepair is to select appropriate $S_{m}$ and $S_{r}$ for each batch, and determine the values of the decision variables $\varphi_{i,j}$, $\phi_{i,j}$, and $\psi_{j}$, such that more STF blocks can be repaired within a shorter time for each batch, as formulated in Eq. \ref{eq_fml_obj}.

\begin{equation}
    \max \frac{|S_{m}|+|S_{r}|}{t_{bch}}
\label{eq_fml_obj}
\end{equation}

\section{Design of IAPRepair}
\label{sec:Design}

IAPRepair is a proactive repair framework for erasure-coded data built on programmable networks. It enhances the parallelism of migration and reconstruction. Specifically, IAPRepair selects STF blocks to be reconstructed for each batch based on the available bandwidth of each node, and assigns reconstruction tasks to individual nodes. Subsequently, based on the reconstruction load of each node, it selects an appropriate number of STF blocks for migration and assigns a replacement node for each migrated STF block. Finally, IAPRepair determines which STF blocks should be reconstructed using in-network aggregation.

\subsection{Block selection for reconstruction}
\label{subsec:Design}

Leveraging in-network aggregation can substantially reduce the bandwidth demand of data reconstruction, which eliminates the bandwidth contention between migration and reconstruction. Consequently, migration and reconstruction can be considered separately. IAPRepair first selects STF blocks to be reconstructed for a batch solely from the perspective of maximizing reconstruction efficiency.


As illustrated by the blue lines in Fig. \ref{fig_Proactive}, since in-network aggregation reduces the volume of data received by nodes during reconstruction, the uplink bandwidth is more likely to become the bottleneck limiting reconstruction efficiency. IAPRepair prioritizes the utilization of the uplink bandwidth of healthy nodes.

For any two nodes $h_{i}$ and $h_{j}$, if $n_{i}^{u}/u_{i}<n_{j}^{u}/u_{j}$, $h_{i}$ is relatively more idle than $h_{j}$. When adding reconstruction blocks to a batch, priority should be given to assigning reconstruction tasks to idle nodes, so as to improve the utilization of their uplink bandwidth. The notation $K_{min}(s_{j})$ denotes the $k$ nodes in $pro(s_{j})$ with the smallest ratio of $(n_{i}^{u}+1) / u_{i}$. Eq. \ref{eq_fml_scj} is used to compute the score $sc_{j}$ of  $s_{j}$. By selecting the STF block with the minimum score, IAPRepair preferentially assigns reconstruction tasks to less loaded nodes.

\begin{equation}
    sc_{j} = \sum_{i\in K_{min(s_j)}} \frac{n_{i}^{u}+1}{u_{i}}
\label{eq_fml_scj}
\end{equation}

The number of STF blocks to be reconstructed in each batch is not fixed. It is determined by the balance of transmission loads across nodes. Let $l^{u}_{avg}=\frac{1}{|H_{h}|}\sum_{i\in H_{h}} \frac{n_{i}^{u}}{u_{i}}$ denote the average load of all nodes, and let $l_{max}^{u}=\max_{h_{i}\in H_{h}}\frac{n_{i}^{u}}{u_{i}}$  denote the maximum load. IAPRepair uses the ratio $l_{avg}^{u}/l_{max}^{u}$  to represent the load balance among nodes. When $l_{avg}^{u}/l_{max}^{u}>\xi$, no additional STF blocks are added to $S_{r}$, where $\xi$ is a predefined threshold. IAPRepair limits the number of reconstructed blocks in each batch to at most $2|H_{h}|$ to avoid long $t_{bch}$ and excessively consumption of computation and cache resources. 

Similarly, $n_{i}^{r,d}/d_{i}$ reflects the receive load of of $h_{i}$, where $n_{i}^{r,d}$ denotes the number of blocks received by $h_{i}$ for block reconstruction in a batch. Among all nodes satisfying the fault-tolerance requirements, the one with the minimum receiving load is selected as the replacement node, as shown in Eq. \ref{eq_fml_tgt_sc}. When multiple STF nodes are detected, several STF blocks may belong to the same stripe, and these blocks must be placed on distinct nodes. To this end, we use $P_{i}$ to record the replacement nodes allocated for the $i$-th stripe. $P_{i}$ persists throughout the entire lifecycle of proactive repair and remains valid across all batches. It is initialized as $P_{g(s_{j})}=\emptyset,\forall j\in S$. Since the replacement node is required to receive $k$ blocks, the numerator in the receiving load computation is incremented by $k$. For reconstruction, the maximum receiving load is: $l_{max}^{r,d}=\max_{h_{i}\in H_{h}}\frac{n_{i}^{r,d}}{d_{i}}$.  

\begin{equation}
    i^{*}=\argmin_{i\in H_{h}\setminus (pro(s_{j})\cup P_{g(s_{j})})} \frac{n_{i}^{r,d}+k}{d_{i}}
\label{eq_fml_tgt_sc}
\end{equation}


\begin{algorithm}
\caption{Block selection for reconstruction}
\label{alg_recon}
\SetKwData{Or}{\textbf{or}}
\DontPrintSemicolon
\KwIn {STF blolks $S$}
\KwOut {Reconstruction blocks $S_{r}$}
    Set $S_{r}=\emptyset$;\;
    Set $n_{i}^{u}=0$ and $n_{i}^{r,d}=0$, $\forall i\in H_{h}$;\;
    \While{$S \neq \emptyset$ and $|S_{r}|\leq 2|H_{h}|$}{
        Set $sc_{temp}=inf $;\;
        \For {each $s_{j}$ in $S$}{
            Compute $sc_{j}$ using Eq. \ref{eq_fml_scj};\;
            \If{$sc_{j} < sc_{temp}$}{
                Set $sc_{temp}=sc_{j}$;\;
                Set $H_{temp}=K_{min}(s_{j})$;\;
                Set $j^{*}=j$;\;
            }
        }
        Select replacement node $h_{i^{*}}$ for $s_{j^{*}}$ using Eq. \ref{eq_fml_tgt_sc};\;
        Add $i^{*}$ to $P_{g(s_{temp})}$, and add $s_{j^{*}}$ to $S_{r}$;\;
        Set $\varphi_{i,j^{*}} = 1$ and $n_{i}^{u} = n_{i}^{u} + 1$ ,$\forall i \in H_{temp}$;\;
        Set $\phi_{i^{*}, j^{*}} = 1$ and $n_{i^{*}}^{d} = n_{i^*}^{d} + k$;\;
        Remove $s_{j^{*}}$ from $S$;\;
        \If{$\frac{l_{avg}^{u}}{l_{max}^{u}}\geq \xi$}{
            break;\;
        }
    }
    return $S_{r}$;\;
\end{algorithm}

Algorithm \ref{alg_recon} describes the details of selecting STF blocks for reconstruction. The input $S$ represents all blocks in the STF nodes, and the output $S_{r}$ denotes the selected STF blocks. Algorithm \ref{alg_recon} performs multiple iterations over $S$. In each iteration, it selects the STF block $s_{temp}$  with the minimum score, along with the set of $k$ nodes $H_{temp}$ responsible for providing data for reconstructing $s_{temp}$ (Lines 4–10). Select the replacement node for $s_{temp}$, and update $P_{g(s_{j})}$ accordingly (Lines 11-12). Identify the data-providing nodes for reconstructing $s_j^{*}$ and update their loads (Line 13). Identify the replacement node for $s_j^{*}$ and update its load (Line 14). Subsequently, remove $s_{j^{*}}$ from $S$ before proceeding to select the next STF block (Line 15). The selection of reconstruction blocks is completed when the ratio of $l_{avg}^{u}$ to $l_{max}^{u}$ exceeds the threshold $\xi$ (Lines 16-17).

\begin{figure}[ht]
\centering
\includegraphics[width=0.85\columnwidth]{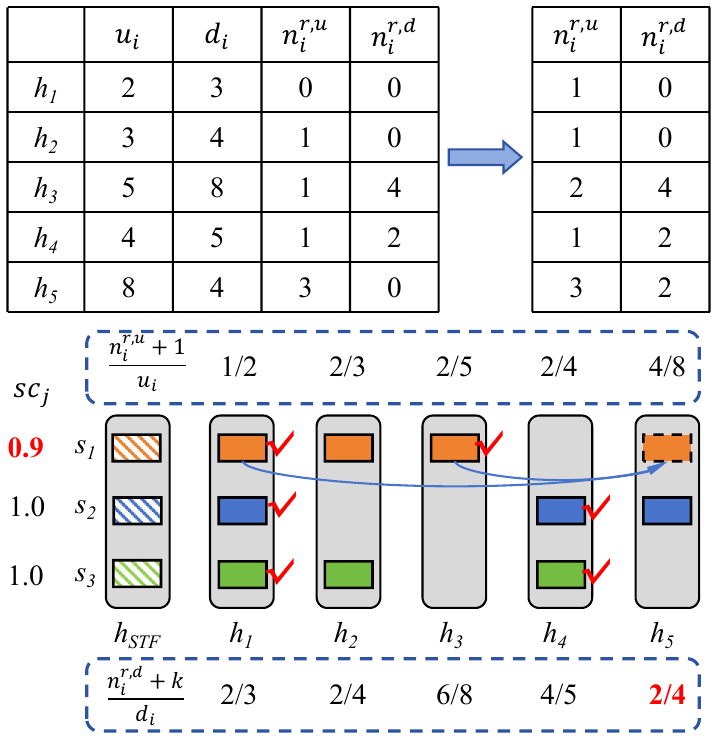}
\caption{Example of block selection for reconstruction}
\label{fig_alg1_demo}
\end{figure}

Fig. \ref{fig_alg1_demo} illustrates an example of Algorithm \ref{alg_recon}, which considers one STF node and five healthy nodes with $RS(4,2)$ encoding. Suppose three STF blocks have already been selected for reconstruction in the batch. Based on Eq. \ref{eq_fml_scj}, the scores $sc_{1}$, $sc_{2}$, and $sc_{3}$ are computed to be 0.9, 1.0, and 1.0, respectively. The block $s_{1}$ with the minimum score is selected and added to the batch, with $h_{1}$ and $h_{3}$ designated as the provider nodes for reconstructing $s_{1}$. Since $h_{1}$ and $h_{2}$, and $h_{3}$ already host blocks belonging to the same stripe as $s_{1}$, the reconstructed block can only be placed on either $h_{4}$ or $h_{5}$. The more idle node $h_{5}$ is selected as the replacement node. Finally, the values of $n_{1}^{u}$, $n_{3}^{u}$, and $n_{5}^{r,d}$ are updated.

\subsection{Block selection for migration}

In this section, we identify the migration blocks. The time required to reconstruct the STF blocks in each batch is primarily determined by $l_{max}^{u}$ and $l_{max}^{r,d}$. To prevent the repair time of the entire batch from increasing due to excessive STF block migration, the maximum number of blocks transmitted by each STF node can be computed using Eq. \ref{eq_stf_max_u}. With in-network aggregation employed, $h_{i}$ is required to receive at least $n_{i}^{r,d}/k $ blocks for reconstruction. Consequently, the maximum number of migration blocks each healthy node can receive is computed by Eq \ref{eq_stf_max_d}. The number of migration blocks is constrained by $N_{i}^{u}$ and $N_{i}^{m,d}$.

\begin{equation}
    N_{i}^{u}=\left\lfloor u_{i}\max (l_{max}^{u},\frac{l_{max}^{r,d}}{k}) \right\rfloor,\forall i\in H_{STF}
\label{eq_stf_max_u}
\end{equation}

\begin{equation}
    N_{i}^{m,d}=\left\lfloor d_{i}\max (l_{max}^{u},\frac{l_{max}^{r,d}}{k})-\frac{n_{i}^{r,d}}{k} \right\rfloor,\forall i\in H_{h}
\label{eq_stf_max_d}
\end{equation}

The node that provides the data for each migration block is uniquely determined, whereas its replacement node is not. Let $H^{m,d}=\{i|i\in H_{h},N_{i}^{m,d}>0\}$  denote the set of healthy nodes capable of receiving migration blocks. For a migration block $s_{j}$, any node in $(H_{h}\setminus pro(s_{j}))\cap H^{m,d}$ can serve as its replacement node. When selecting replacement nodes for migration blocks, concentrating these blocks on a small number of healthy nodes is beneficial for reducing the resource consumption of the programmable switch. This is because the bandwidth released by reconstructing one block via in-network aggregation can be leveraged to migrate $k-1$ blocks. Distributing migrated blocks across different nodes would lead to more idle bandwidth and require reconstructing more STF blocks using in-network aggregation. 

\begin{figure}[ht]
\centering
\includegraphics[width=0.60\columnwidth]{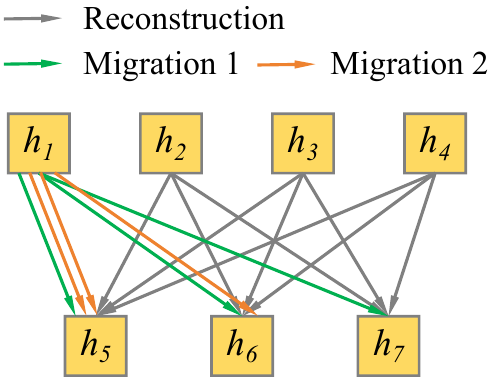}
\caption{Example of replacement node selection for migration}
\label{fig_alg2_demo}
\end{figure}

As illustrated in Fig. \ref{fig_alg2_demo}, $h_{1}$ is the STF node, and all the others are healthy nodes. In this batch, three blocks are reconstructed and three blocks are migrated concurrently. To prevent the receiving side from becoming a bottleneck that degrades repair efficiency, the number of blocks received by any single node must not exceed three. If the migration blocks are placed according to the scheme indicated by the green lines, in-network aggregation is required to reconstruct all three blocks. In contrast, adopting the scheme indicated by the orange lines, only two blocks need to be reconstructed through in-network aggregation. For all migration blocks, IAPRepair preferentially selects the node with the largest $N_{i}^{m,d}$ as their replacement nodes.

\begin{algorithm}
\caption{Block selection for migration}
\label{alg_mgrt}
\SetKwData{Or}{\textbf{or}}
\DontPrintSemicolon
\KwIn {STF blolks $S$}
\KwOut {Migration blocks $S_{m}$}
    Set $Q_{i}=\emptyset,\forall H_{h}$;\;
    Set $n_{i}^{u}=0,\forall i\in H_{STF}$;\;
    Set $n_{i}^{m,d}=0,\forall i\in H_{h}$;\;
    Compute $N_{i}^{u}$ and $N_{i}^{m,d}$ using Eqs. \ref{eq_stf_max_u} and \ref{eq_stf_max_d}, respectively;\;
    \For{each $s_{j}$ in $S$}{
        Set $i^{*}= \argmin_{i\in H_{h}\setminus (pro(s_{j})\cup P_{g(s_{j})})} N_{i}^{m,d}$;\;
        Add $s_{j}$ to $Q_{i^{*}}$ and add $i^{*}$ to $P_{g(s_{j})}$;\;
    }
    \For{each $Q_{l}$ in $sort(Q_{1},Q_{2},\dots,Q_{|H_{h}|})$}{
        \For{each $s_{j}$ in $Q_{l}$}{
            Let $h_{i}$ be the STF node of $s_{j}$;\;
            \If{$n_{i}^{u}<N_{i}^{u}$ and $n_{i^{*}}^{m,d}<N_{i^{*}}^{m,d}$}{
                Set $n_{i}^{u}=n_{i}^{u}+1$ and $n_{i^{*}}^{m,d}=n_{i^{*}}^{m,d}+1$;\;
                Add $s_{j}$ to $S_{m}$;\;
                Set $\varphi_{i,j} = 1$ and $\phi_{i^{*},j} = 1$;\;
                Remove $s_{j}$ from $S$;\;
            }
        }
    }
    return $S_{m}$;\;
\end{algorithm}

Algorithm \ref{alg_mgrt} presents the details of migration task assignment. $Q_{i}$ stores all migrated STF blocks that designate $h_{i}$ as their replacement node. $Q_{i}$ is initialized as empty (Line 1). The number of migration blocks to be transmitted by each STF node ($n_{i}^{u},\forall i\in H_{STF}$ ) and the number of migration blocks to be received by each healthy node ($n_{i}^{m,d},\forall i\in H_{h}$) are both initialized to 0 (Lines 2–3). The values of $N_{i}^{u}$ and $N_{i}^{m,d}$ are computed using Eqs. \ref{eq_stf_max_u} and \ref{eq_stf_max_d}, respectively (Line 4). While ensuring fault tolerance, for each STF block $s_{j}$ in $S$, the node $h_{i^{*}}$ with the largest $N_{i}^{m,d}$ is selected as its replacement node and recorded in $Q_{i^{*}}$ (Lines 5–7). Traverse $\{Q_{1},Q_{2},\dots,Q_{|H_{h}|}\}$ in descending order according to the number of migration blocks contained in each set (Lines 8-9). For an STF block $s_{j}$, if the number of blocks transmitted by its assigned STF node $h_{i}$ is less than $N_{i}^{u}$, and the number of blocks received by its assigned replacement node $h_{i^{*}}$ is less than $N_{i^{*}}^{m,d}$, then $s_{j}$ can be migrated in the current batch (Line 11). Accordingly, $s_{j}$ is added to $S_{m}$, and $\varphi_{i,j}$ and $\phi_{i^{*},j}$ are set to 1 (Lines 12-15).

\subsection{In-network aggregation decision}

IAPRepair traverses each healthy replacement node to identify which of its assigned reconstruction blocks should be repaired via in-network aggregation. Let $n_{i}^{ina}$ denote the number of reconstruction blocks that require repair via in-network aggregation and are assigned $h_{i}$ as the replacement node. The value of $n_{i}^{ina}$ can be computed according to Eq. \ref{eq_subsec3_1}. To avoid an increase in repair time due to insufficient data receiving capacity, $n_{i}^{ina}$  must satisfy the constraint given in Eq. \ref{eq_subsec3_2}. By transforming Eq. \ref{eq_subsec3_2}, it can be derived that when $n_{i}^{ina}$ at least equals to $N_{i}^{ina}$ (Eq. \ref{eq_subsec3_3}), the limitation on repair performance imposed by receiving capacity is eliminated.

\begin{equation}
    n_{i}^{ina}=\sum_{j\in S_{r}}\phi_{i,j}\psi_{j}, \forall i\in H_{h}
\label{eq_subsec3_1}
\end{equation}

\begin{equation}
    \frac{n_{i}^{m,d}}{d_{i}}+\frac{k(n_{i}^{r,d}-n_{i}^{ina})}{d_{i}}+\frac{n_{i}^{ina}}{d_{i}}\leq l_{max}^{u},\forall i\in H_{h}
\label{eq_subsec3_2}
\end{equation}


\begin{equation}
    N_{i}^{ina} = \max \left( \left\lceil \frac{d_{i}l_{max}^{u}-n_{i}^{m,d}-kn_{i}^{r,d}}{1-k} \right\rceil, 0 \right), \forall i \in H_{h}
\label{eq_subsec3_3}
\end{equation}

However, the value of $n_{i}^{ina}$ is also constrained by the internal resources of the programmable switch  (Eq. \ref{eq_fml_psi_sum}). Thus it cannot be guaranteed that setting $n_{i}^{ina}$ to $N_{i}^{ina}$ is always feasible. IAPRepair mitigates the impact of the internal resource constraints of the programmable switch on repair performance by minimizing $\max_{i\in H_{h}}(N_{i}^{ina}-n_{i}^{ina})$, where $N_{i}^{ina}\geq n_{i}^{ina}$. Algorithm \ref{alg_ina} details the procedure for determining the blocks which are constructed via in-network aggregation. Initially, all reconstructions are first assumed to not require in-network aggregation (Line 1). The demand of each node for in-network aggregation is computed according to Eq. \ref{eq_subsec3_3} (Line 2). $S_{ina}$ and $cnt$ are employed to record the blocks adopting in-network aggregation and the quantity, respectively. In each iteration, Algorithm \ref{alg_ina} identifies the node $h_{i^{*}}$ with the largest difference between $N_{i}^{ina}$ and $n_{i}^{ina}$, and randomly selects one block $s_{j^{*}}$ from all STF blocks whose replacement node is $h_{i^{*}}$ to reconstruct via in-network aggregation  (Lines 6-9). Algorithm \ref{alg_ina} continues until the requirements of all nodes are satisfied or no additional reconstructed blocks can adopt in-network aggregation.

\begin{algorithm}
\caption{In-network aggregation descision}
\label{alg_ina}
\SetKwData{Or}{\textbf{or}}
\DontPrintSemicolon
\KwIn {Limitations on in-network aggregation $\theta$}
\KwOut {Blocks reconstructed via in-network aggregation $S_{ina}$}
    Set $n_{i}^{ina} = 0,\forall i\in H_{h}$;\;
    Compute $N_{i}^{ina},\forall i \in H_{h}$ using Eq. \ref{eq_subsec3_3};\;
    Set $i^{*}=arg \max_{i\in H_{h}}(N_{i}^{ina}-n_{i}^{ina})$;\;
    Set $cnt=0$ and $S_{ina}=\emptyset$;\;
    \While{$cnt<\theta$ and $N_{i^{*}}^{ina}>n_{i^{*}}^{ina}$}{
        Set $cnt=cnt+1$;\;
        Set $n_{i^{*}}^{ina}=n_{i^{*}}^{ina}+1$;\;
        Select a block $s_{j^{*}}$ assigned $h_{i^{*}}$ as the replacement node and add it to $S_{ina}$;\;
        Set $\psi_{j^{*}} = 1$;\;
        Set $i^{*}=\arg\max_{i\in H_{h}}(N_{i}^{ina}-n_{i}^{ina})$;\;   
    }
    return $S_{ina}$;\;
\end{algorithm}

\subsection{Algorithm Analysis}

Complexity Analysis. Algorithm \ref{alg_recon} selects at most $2|H_{h}|$ reconstruction blocks for each batch and traverses $S$ at most $2|H_{h}|$ times. The time complexity of computing the score for one block is O($|H_{h}|$) . Therefore, the time complexity of Algorithm \ref{alg_recon} is O($2|H_{h}|^{2}|S|$) . Algorithm \ref{alg_mgrt} traverses $S$ twice. In the first traversal, a replacement node is selected for each STF block, where each selection incurs a time complexity of $O(|H_{h}|)$. A total of $|S|$ selections are required. The second traversal selects migrated blocks and inserts them into $|S_{m}|$. Thus, the time complexity of Algorithm 2 is $O(|S|(|H_{h}|+1))$ . Algorithm \ref{alg_ina} iterates at most $\theta$ times, and each iteration selects one block to add to $S_{ina}$ with a time complexity of $O(|H_{h}|)$. Therefore, the time complexity of Algorithm \ref{alg_ina} is $O(\theta|H_{h}|)$. Since Algorithms \ref{alg_recon}, \ref{alg_mgrt}, and \ref{alg_ina} execute sequentially, the overall time complexity of IAPRepair is approximately $O(2|H_{h}|^{2}|S|+\theta|H_{h}|)$.

Extensibility. IAPRepair is equally applicable to LRC \cite{HuangSXOCG0Y12,TamoB14}, which is another erasure code deployed in production environments, denoted as $LRC(n,k,r )$. An $LRC(n,k,r )$ stripe comprises $n$ blocks, including $k$ data blocks. These $k$ data blocks are partitioned into $k/r$ groups ($k$ is divisible by $r$ ). Each group contains $r$ data blocks and one local parity block. The local parity block is generated from the data blocks within the same group. In addition, $LRC(n,k,r)$ contains $n-k-k/r$ global parity blocks. These global parity blocks are encoded from all $k$ data blocks. Reconstructing a local block within a group requires only $r$  blocks from the same group, whereas reconstructing a global parity block requires reading $r$ blocks from every group. STF nodes store both global parity blocks and local blocks, and the number of global parity blocks is smaller. Since the data volume required to reconstruct a local block is substantially less than that required for a global parity block, IAPRepair prioritizes repairing global parity blocks through migration.

\section{Implementation}
\label{sec:implementation}

We have implement a prototype of IAPRepair using C++, Python, and the P4 language. The transmission of repair data and repair commands leverages Linux socket interfaces, while the computation of decoding coefficients and Galois field operations are built upon the Jerasure library \cite{plank2008jerasure}. Algorithms \ref{alg_recon}-\ref{alg_ina} of IAPRepair are implemented in Python, and the internal processing logic of the programmable switch is realized in P4. Paint \cite{xia2024parallelized} proposed the ACP protocol for data transmission for in-network aggregation. The ACP protocol extends UDP with a retransmission mechanism to ensure reliable delivery, and enables resource management on programmable switches through additional fields including Ver, Bid, and Pid. Our prototype system adopts the ACP protocol for in-network aggregation.

\begin{figure}[t]
\centering
\includegraphics[width=0.99\columnwidth]{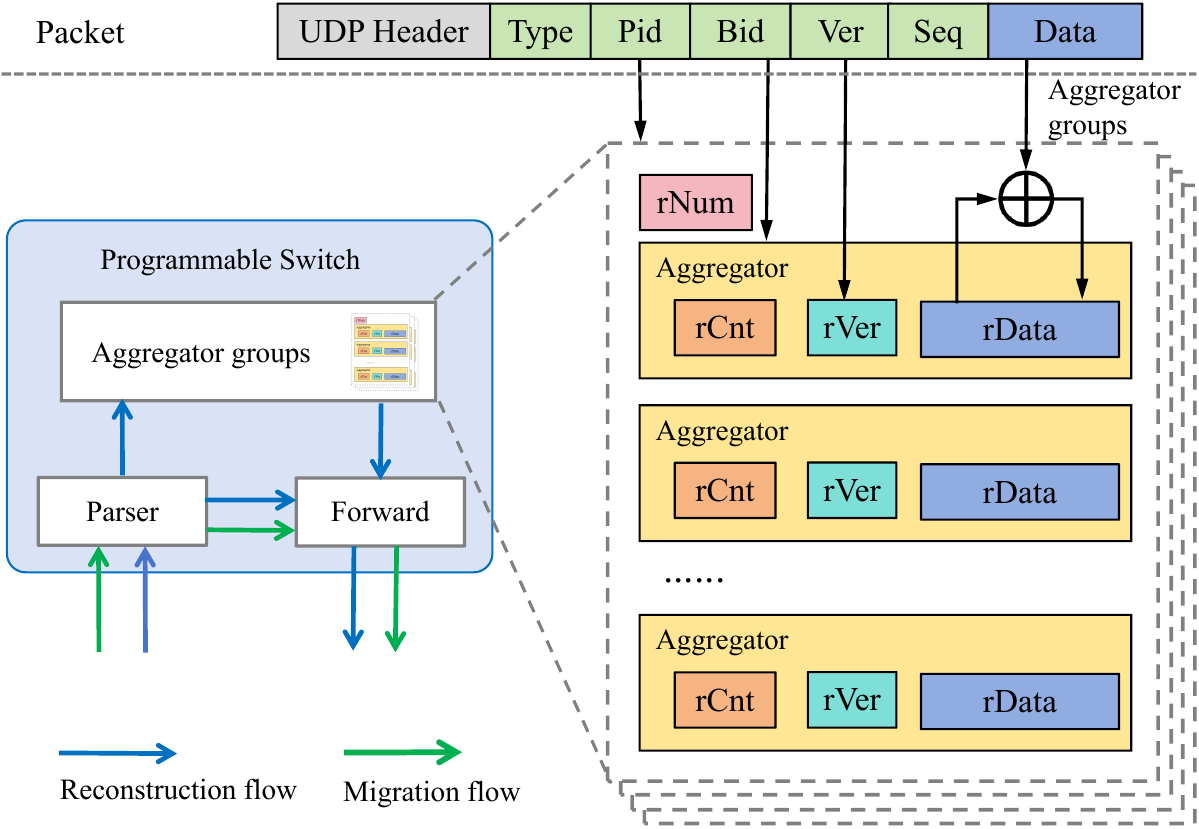}
\caption{The system architecture of IAPRepair.}
\label{fig_implement}
\end{figure}

In the prototype system, the programmable switch is employed for data forwarding and in-network aggregation, as illustrated in Fig. \ref{fig_implement}. The programmable switch comprises the Parser, the Forward module, and multiple aggregators. The Parser identifies whether a packet requires in-network aggregation based on its Type field. Reconstruction packets carrying specific markings are distributed to the aggregator group, while all other packets (including both reconstruction and migration packets) are directly dispatched to the Forward module. The Aggregator serves as the core component for in-network aggregation, and perform aggregation operations on the marked reconstruction packets. The aggregated results are subsequently delivered to the Forward module. In the figure, blue lines represent reconstruction data flows, and green lines represent migration data flows. Only a portion of the reconstruction data requires in-network aggregation. STF nodes transmit migration data, and do not receive any repair data. Healthy nodes transmit reconstruction data and receive both migration and reconstruction data.

IAPRepair allocates a group of aggregators to each STF block reconstructed via in-network aggregation. This design enables nodes to transmit data continuously without awaiting receiver acknowledgments, thereby preventing throughput degradation caused by acknowledgment waiting. Assuming that each group contains $N_{a}$ aggregators, the sender can transmit at least $N_{a}$ packets consecutively. Additionally, multiple aggregators function as data buffers, mitigating the impact of differences in packet arrival times from different nodes on aggregation efficiency.

Each aggregator maintains three state objects: rCnt, rVer, and rData. Aggregators within the same group share a common rNum state object. All of these state objects are implemented using Register objects in the programmable switch. The rNum register stores the number of data flows required for aggregation and is configured before proactive repair begins. The rData register stores the intermediate aggregation result. Upon receiving a packet, the aggregator performs a bitwise XOR operation between rData and the packet’s Data field, writes the result back to rData, and increments rCnt. When the value of rCnt equals to that of rNum, indicating that all aggregation operations required for decoding have been completed, the aggregator dispatches the data stored in rData to the Forward module and resets rData. 

The Pid and Bid fields are used to identify the aggregator group and the specific aggregator within that group, respectively. Packets for repairing the same STF block carry an identical Pid. The Seq field indicates the offset of the carried data within the STF block and is employed to handle out-of-order packet arrivals. Due to the limited resources of programmable switches, a single aggregator is shared by packets with different Seq, with Bid cycling repeatedly from 1 to $N_{a}$. However, aggregator sharing may lead to aggregation of packets with differing Seq. To prevent this, the sender increments the Ver field by 1 after every $N_{a}$  packets transmitted. Correspondingly, rVer is incremented by 1 upon completing aggregation of rNum packets. If the value of Ver differs from that of rVer, a transmission error is detected, and the corresponding data must be retransmitted.

\section{Evaluation}
\label{sec:evaluation}

We conduct a complete proactive repair process on the IAPRepair prototype system to evaluate its repair performance. We also evaluate the resource consumption and runtime overhead of IAPRepair in large-scale storage systems via simulation. In our evaluation, the repair time includes the data transfer time and the algorithm runtime.

\subsection{Experimental settings}
The prototype system consists of a Tofino ASIC programmable switch (with bf-sde 9.2.0 Software Development Kit installed) and 17 servers. Among these servers, one functions as the management node, and the remaining 16 operate as storage nodes. Each server is equipped with an Intel Core i7-4790K CPU, 16 GB of RAM, a 500 GB SSD, and a dual-port 10GbE network interface card, and has the Ubuntu 20.04 operating system installed.

Unless otherwise specified, all the results are the average values of 10 tests. The default coding is $RS(9,6)$, the block size is 8 MB, and the value of $\xi$ is set to 0.9. There are 512 stripes and 1 STF node. Each stripe consists of one block placed in the STF node, while the other $n-1$ blocks are randomly placed across healthy nodes. Recent work \cite{yao2022pivotrepair} evaluates the bandwidth consumption of each node in a storage system under three workloads: SWIM, TPC-DS, and TPC-H. We randomly select measurement results exhibiting heterogeneous bandwidth consumption and use the Linux tc command to configure the available bandwidth of each node in our prototype system accordingly. 

We select FastPR \cite{li2022fast}, SelectiveEC \cite{xu2021selectiveec} and Migration as the comparative algorithms. FastPR parallelizes migration and reconstruction to proactively repair blocks in STF nodes, but its block selection strategy does not adequately account for load balancing across nodes. SelectiveEC proposes a load-balanced task assignment method, but it only adopts reconstruction for repair. Migration directly transfers data from STF nodes to healthy nodes without performing data reconstruction.  

\subsection{Repair time}

\begin{figure}[t]
\centering
\includegraphics[width=.72\columnwidth]{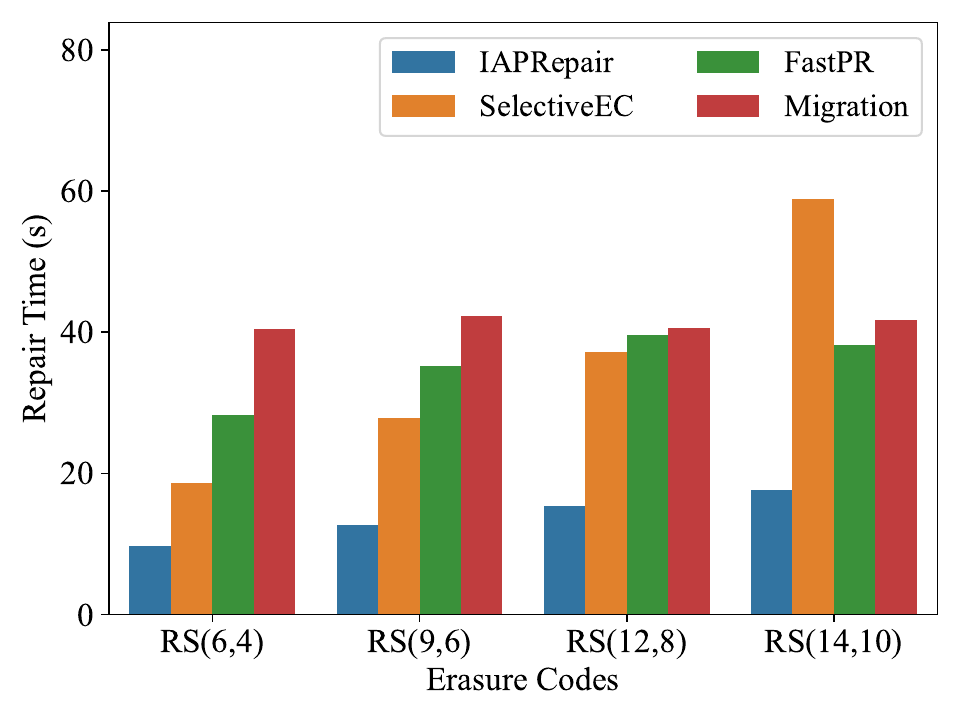}
\caption{Impact of coding parameters on repair time}
\label{fig_exp01}
\end{figure}

Experiment 1 (Impact of coding parameters): Fig. \ref{fig_exp01} presents the repair time of different methods under the coding parameters of $RS(6, 4)$ \cite{zhang2013accelerate}, $RS(9, 6)$ \cite{ShvachkoKRC10}, $RS(12, 8)$ \cite{lai2015atlas} and $RS(14, 10)$ \cite{muralidhar2014f4}. IAPRepair leverages in-network aggregation to reduce the bandwidth consumption of reconstruction and eliminates the bottleneck imposed by node downlink bandwidth on repair performance. Additionally, the block selection and balanced task assignment methods improve reconstruction efficiency. IAPRepair consistently demonstrates the best repair performance across all coding parameters, reducing repair time by 47.83\% to 68.04\% compared to SelectiveEC, 53.79\% to 65.49\% compared to FastPR, and 57.76\% to 75.94\% compared to Migration. 

As the coding parameter $k$ increases, reconstructing a block requires more data, leading to a corresponding increase in repair time. SelectiveEC exhibits a significant degradation in repair performance in $RS(14,10)$. The primary reason stems from the difficulty in establishing a one-to-one matching between STF stripes and replacement nodes when the parameter $k$ is close to the number of nodes. Although IAPRepair also encounters the issue of shared replacement nodes, it effectively avoids performance degradation by reducing data transmission volume through in-network aggregation. Migration transfers data from the STF node sequentially, resulting in long repair times, yet its repair duration is unaffected by coding parameters.

\begin{figure}[t]
\centering
\includegraphics[width=.72\columnwidth]{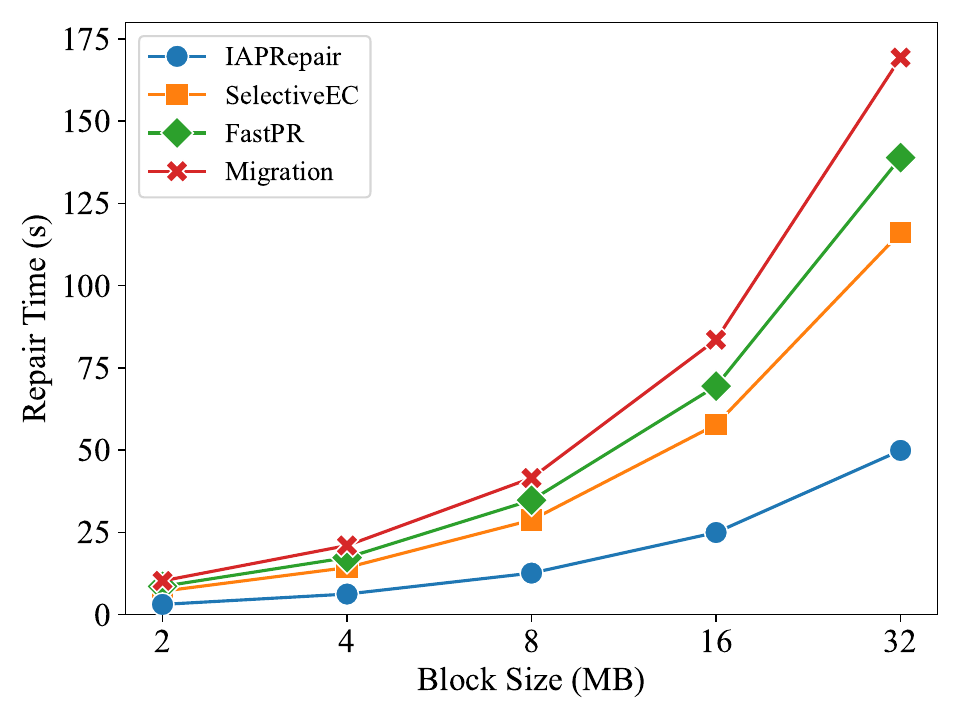}
\caption{Impact of block size on repair time}
\label{fig_exp02}
\end{figure}

Experiment 2 (Impact of block size): We evaluated the impact of block size on repair time, with the block size varying from 2MB to 32MB. The experimental results are illustrated in Fig. \ref{fig_exp02}, showing that the repair time of all methods increases with larger block sizes, yet IAPRepair consistently outperforms other methods. On average, the repair time of IAPRepair is 55.31\%, 61.97\% and 67.87\% shorter than that of SelectiveEC, FastPR and Migration, respectively.

Experiment 3 (Impact of the number of nodes): 
As the number of storage nodes increases, more STF blocks can be reconstructed concurrently. Consequently, the repair time for IAPRepair, SelectiveEC and FastPR decreases accordingly. SelectiveEC incurs long repair times when the number of nodes is close to the coding parameter $k$. FastPR requires an increase of $k$ nodes to add one reconstructed STF stripe in a repair batch, resulting in a step-like pattern in its repair time curve.  Due to the limitation of the uplink bandwidth of the STF node, increasing the number of healthy nodes does not reduce the repair time of Migration. IAPRepair exhibits the shortest repair time, reducing it by at least 40.97\%, 47.82\%, and 60.48\% compared to SelectiveEC, FastPR, and Migration, respectively. In addition, IAPRepair exhibits a smoothly decreasing repair time curve, indicating that it can exploit the bandwidth resources of all nodes to improve repair performance.

\begin{figure}[t]
\centering
\includegraphics[width=.72\columnwidth]{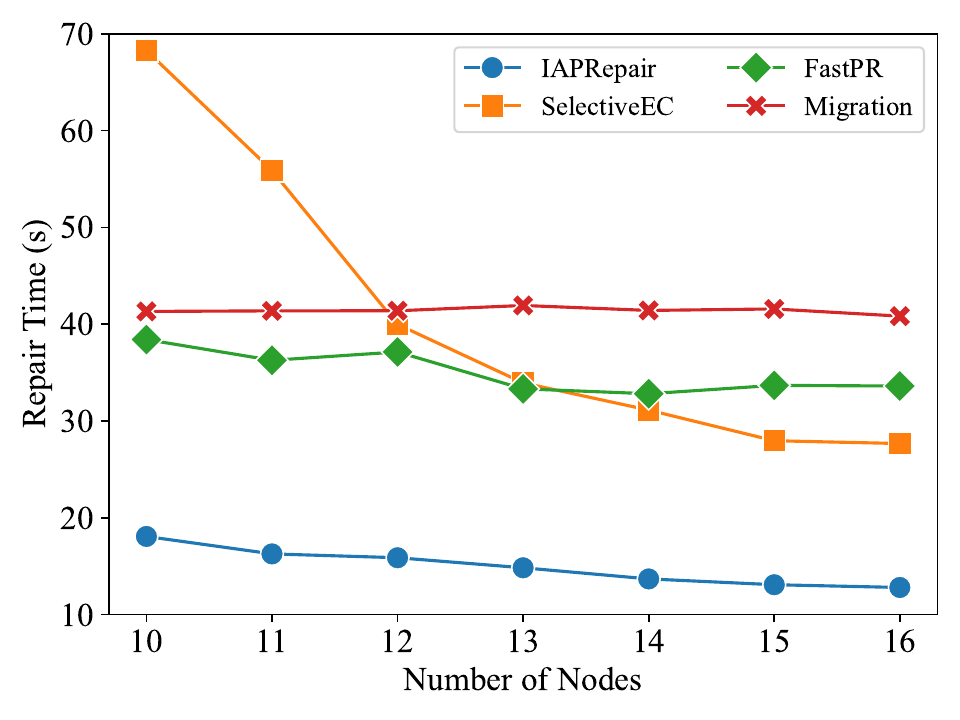}
\caption{Impact of the number of nodes on repair time}
\label{fig_exp03}
\end{figure}

\begin{figure}[t]
\centering
\includegraphics[width=.72\columnwidth]{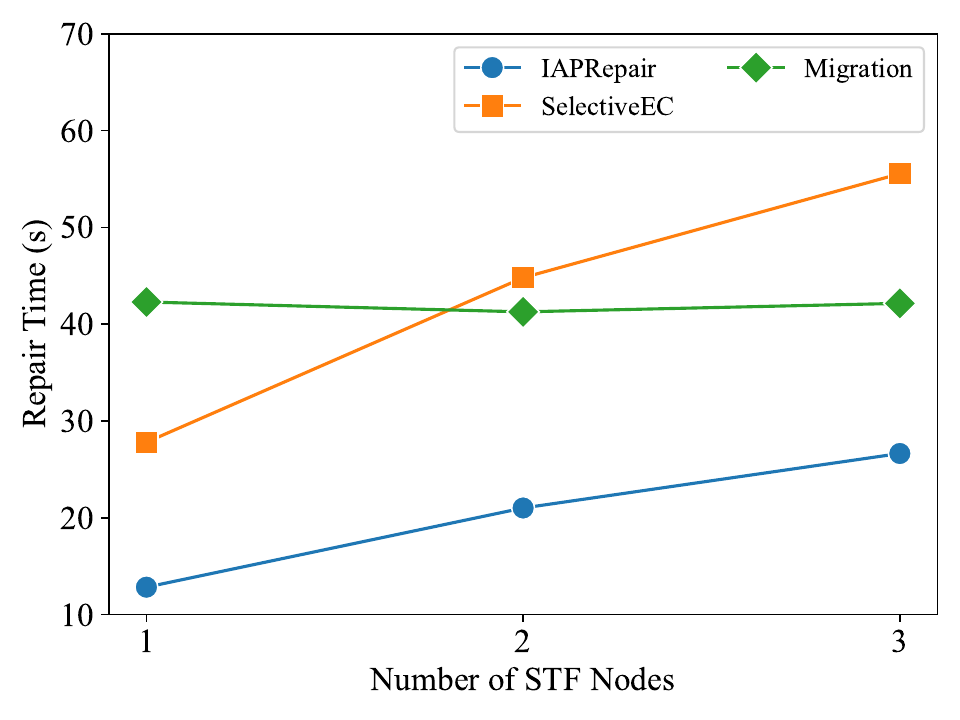}
\caption{Impact of number of STF nodes on repair time}
\label{fig_exp04_nSTF}
\end{figure}

Experiment 4 (Impact of number of STF nodes): 
We evaluate proactive repair for multiple STF nodes. For the $RS(9,6)$ coding scheme used in our experiments, data reconstruction becomes infeasible when more than three nodes fail. Thus, the number of STF nodes ranges from 1 to 3. Since FastPR is not applicable to the repair of multiple STF nodes, only SelectiveEC and Migration were considered as baseline methods in the experiments. As the number of STF nodes increases, the number of healthy nodes available for reconstruction decreases, resulting in degraded reconstruction performance. Because healthy nodes are more than STF nodes, the data stored on multiple STF nodes can be migrated in parallel without mutual interference. SelectiveEC repairs through reconstruction, and its repair time increases noticeably with the number of STF nodes. Migration copies data from STF nodes to healthy nodes, exhibiting nearly constant repair time.  IAPRepair combines migration and reconstruction. Although its repair time also increases as the number of STF nodes grows, it consistently outperforms the other methods. As shown in Fig. 14, compared with the baseline methods, IAPRepair reduces the repair time by 53.84\%, 49.09\%, and 36.79\% in scenarios with one, two, and three STF nodes, respectively.

\begin{figure}[t]
\centering
\includegraphics[width=.72\columnwidth]{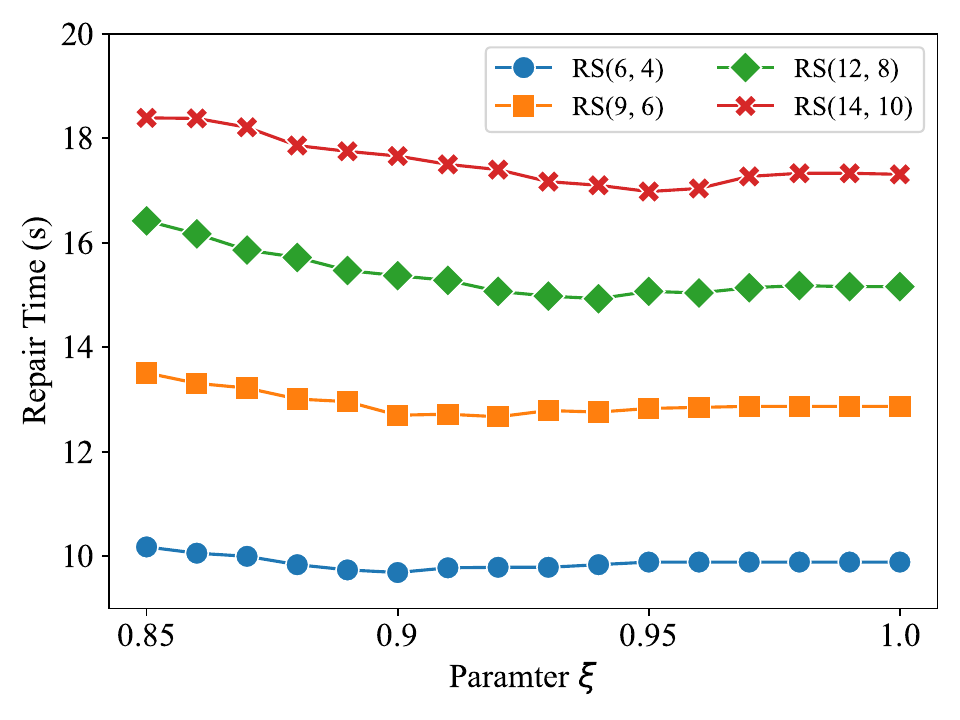}
\caption{Impact of $\xi$ on repair time}
\label{fig_exp05_xi_1}
\end{figure}

\begin{figure}[t]
\centering
\includegraphics[width=.72\columnwidth]{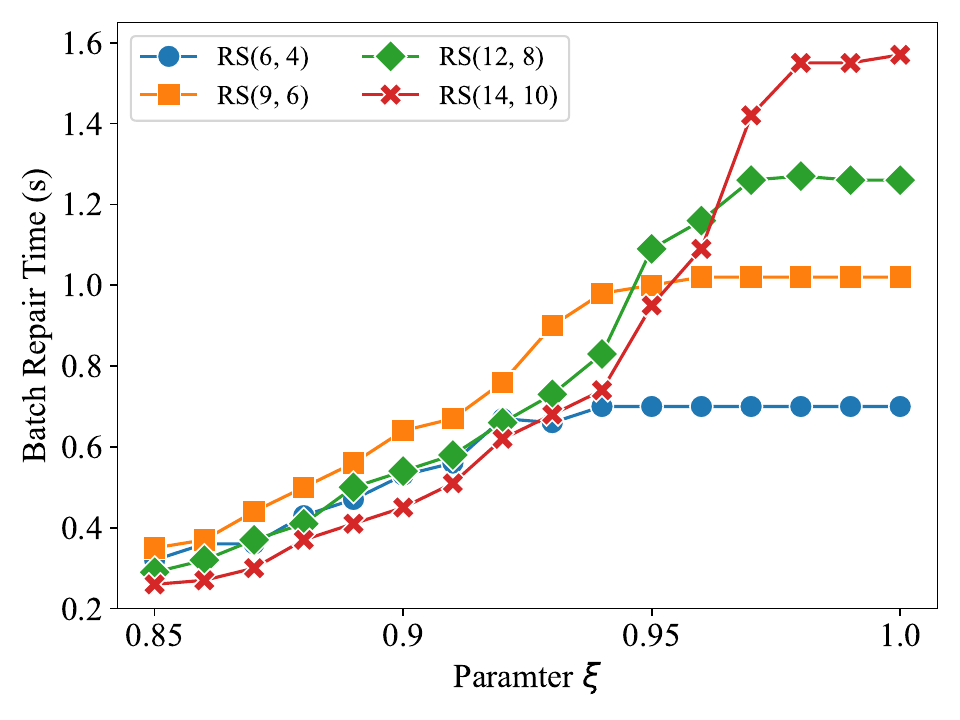}
\caption{Impact of $\xi$ on batch time}
\label{fig_exp05_xi_2}
\end{figure}

Experiment 5 (Impact of $\xi$): We evaluated the impact of parameter $\xi$ on repair time, with $\xi$ ranging from 0.85 to 1.0. As illustrated in Fig. \ref{fig_exp05_xi_1}, the repair time first decreases as the value of parameter $\xi$ increases, and then rises gradually. As $\xi$ increases, the load distribution across nodes becomes more balanced, which is conducive to improving the repair efficiency. However, an excessively high threshold cannot be reached in practice, and Algorithm \ref{alg_recon} terminates after selecting $2|H_{h}|$ STF blocks. At this point, the balance among nodes may deteriorate, resulting in an increase in the repair time. The shortest repair time is achieved when $\xi$ is within the range of 0.9 to 0.95. Fig. \ref{fig_exp05_xi_2} further shows that the repair time of each batch increases with $\xi$, because Algorithm $\ref{alg_recon}$ needs to select more STF blocks to meet the threshold requirement. Since network conditions vary dynamically, an excessively long batch time may cause the network state measured during the execution of Algorithm 1 to become obsolete. Therefore, $\xi$ should not be set too large. In the experiments, $\xi$ is set to 0.9.

\begin{figure}[t]
\centering
\includegraphics[width=.72\columnwidth]{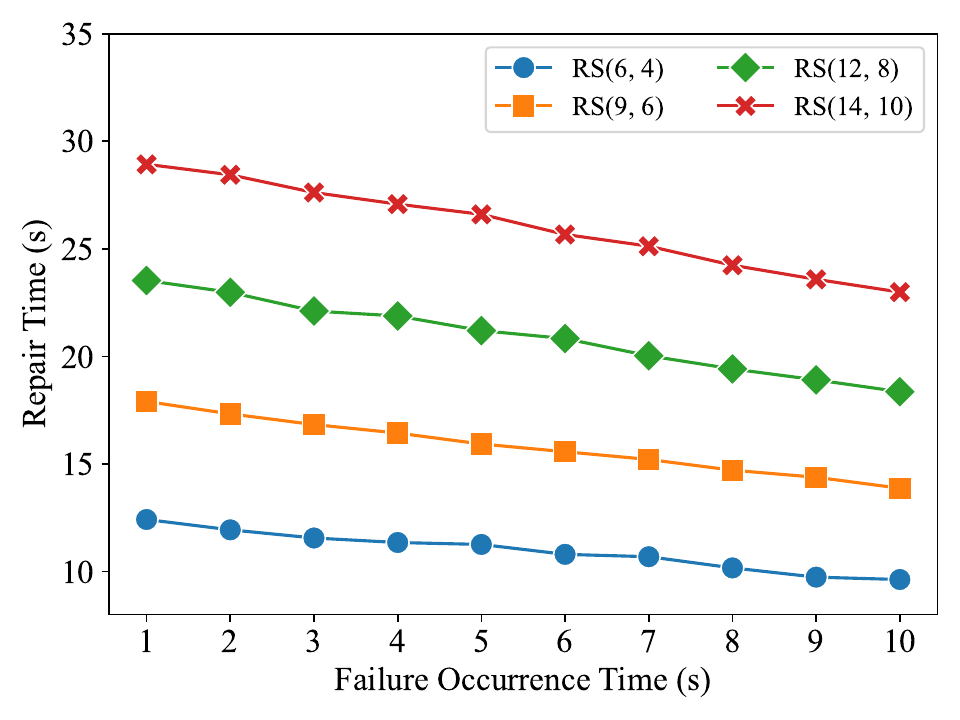}
\caption{Impact of node STF failure time}
\label{fig_exp06}
\end{figure}

Experiment 6 (Impact of node failure): STF nodes may fail during the repair process. To evaluate the impact of STF node failures on repair performance, we configured the STF node to fail at different time points after the proactive repair began. Once an STF node fails, IAPRepair can only perform reconstruction. The test results are presented in Fig. \ref{fig_exp06}. IAPRepair can still accomplish the node repair even the STF node fails. The repair performance exhibits higher sensitivity to the failure occurrence time with a smaller coding parameter $k$. Specifically, for $RS(6,4)$, $RS(9,6)$, $RS(12,8)$, and $RS(14,10)$, a failure occurring at the 10th second reduces the repair time by 22.56\%, 22.31\%, 21.97\%, and 20.53\%, respectively, compared with a failure occurring at the 1st second. 

\begin{figure}[t]
\centering
\includegraphics[width=.72\columnwidth]{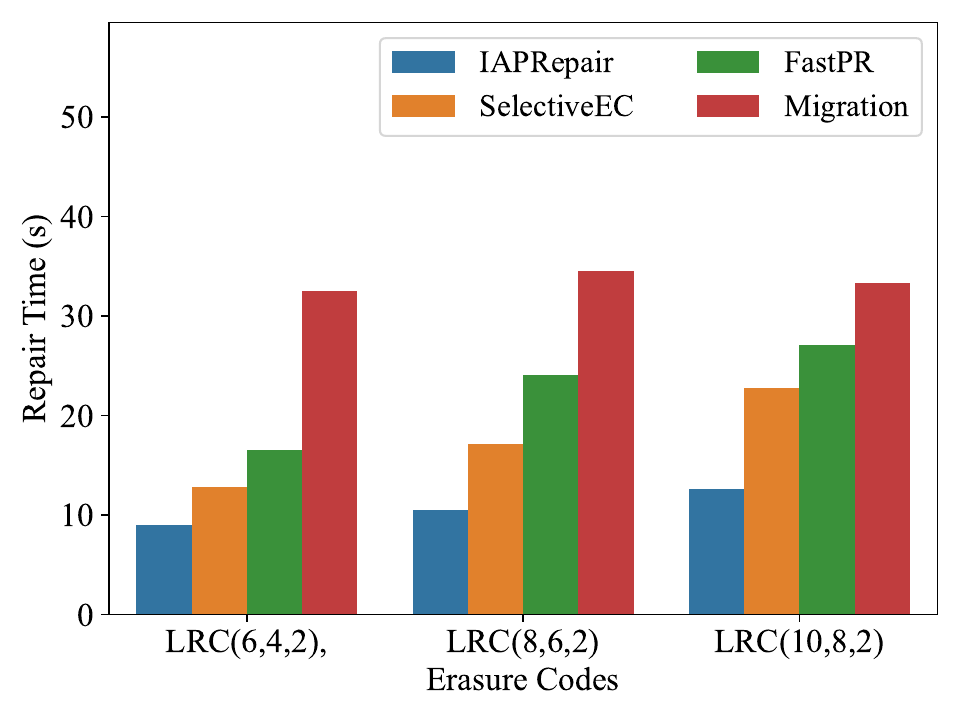}
\caption{Extension to LRCs}
\label{fig_exp07}
\end{figure}

Experiment 7 (Extension to LRCs): We use $LRC(n,k,r)$ to evaluate the generality of IAPRepair. An $LRC(n,k,r)$ consists of data blocks, local parity blocks, and global parity blocks. Since these different types of blocks require different reconstruction procedures, proactive repair in LRCs is more complex. We select $LRC(6,4,2)$, $LRC(8,6,2)$, and $LRC(10,8,2)$ in this evaluation. he evaluation results are presented in Fig. \ref{fig_exp07}. Compared with the baseline methods, IAPRepair reduces the repair time by 29.79\%–69.65\%. These results demonstrate that IAPRepair can be applied to various erasure codes.

\begin{figure}[t]
\centering
\includegraphics[width=.72\columnwidth]{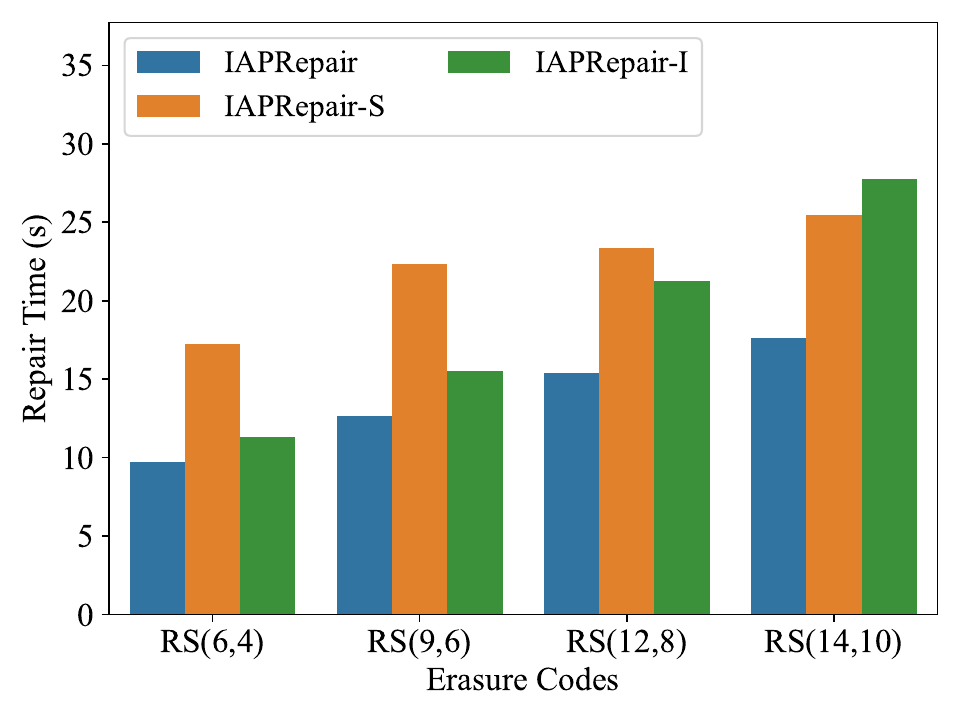}
\caption{Ablation study}
\label{fig_exp08}
\end{figure}

Experiment 8 (Ablation study on IAPRepair): 

We perform an ablation study on IAPRepair to investigate the effectiveness of block selection and in-network aggreation decision. IAPRepair-S and IAPRepair-I are two variants of IAPRepair. IAPRepair-S randomly select STF blocks for reconstruction. IAPRepair-I performs data aggregation within nodes. The experimental results are shown in Fig. \ref{fig_exp08}. When the number of nodes is comparable to the coding parameter, the block selection algorithm is more effective in reducing repair time than in-network aggregation. Conversely, when the number of nodes is markedly larger than the coding parameters, in-network aggregation becomes more effective. IAPRepair is the synthesis of IAPRepair-S and IAPRepair-I, and achieves the shortest repair time. Compared with IAPRepair-S, IAPRepair reduces the repair time by up to 43.47\%; compared with IAPRepair-I, it achieves a maximum reduction of 36.42\% in repair time.

\subsection{Repair Overhead}

\begin{figure}[t]
\centering
\includegraphics[width=.75\linewidth]{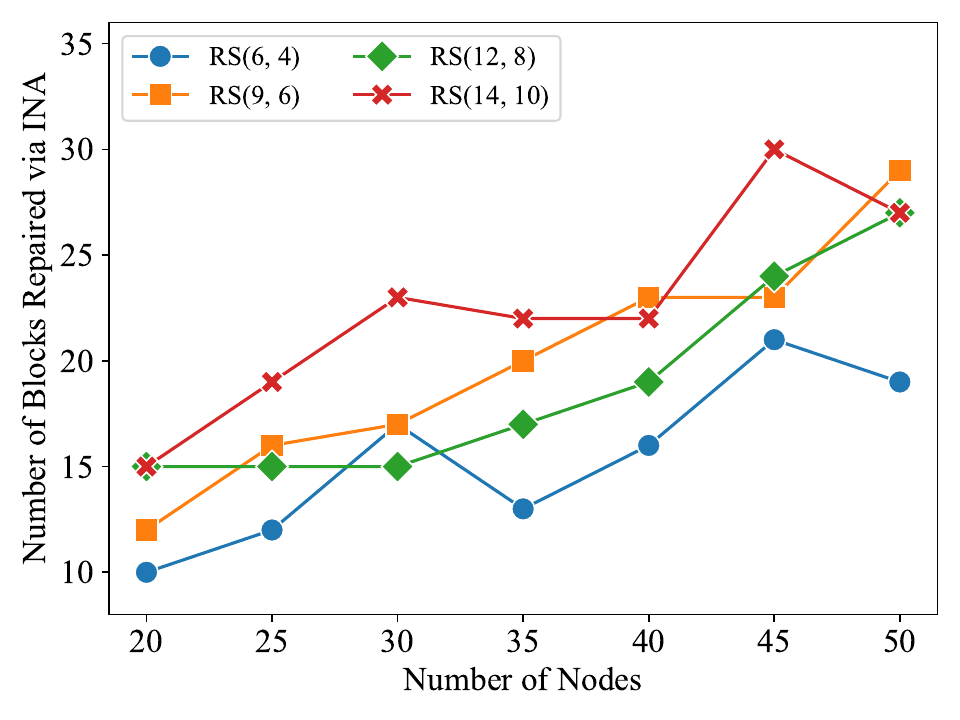}
\caption{Number of blocks reconstructed using INA} 
\label{fig_exp09}
\end{figure}

Experiment 9 (Resource usage of the switch): We evaluate the internal resource usage of the programmable switch. The resource usage is directly proportional to the number of blocks reconstructed using in-network aggregation. We record the maximum number of STF blocks that are reconstructed via in-network aggregation across all batches under various coding parameters and number of nodes. As illustrated in Fig. \ref{fig_exp09}, the number of nodes increases from 20 to 50. The number of STF blocks reconstructed via in-network aggregation exhibits an upward trend as the number of nodes increases. For $RS(6,4)$, $RS(9,6)$, $RS(12,8)$ and $RS(14,10)$, the maximum number of blocks requiring in-network aggregation is 21, 29, 27 and 30, respectively. We establish 32 aggregator groups in the programmable switch, each containing 64 aggregators. The payload size of each packet is 400 bytes. In-network aggregation primarily consumes SRAM resources in the programmable switch, and the corresponding resource consumption is presented in Table \ref{tab_exp09}. The in-network aggregation operations are distributed across 11 stages, with the SRAM utilization in each stage remaining below 15.00\%. The usage of Hash Bit and VLIW Instruction resources indicates that each stage retains sufficient capacity for resource mapping and instruction execution. TCAM is primarily used in packet-forwarding operations for rapid address matching, mask-based matching, and longest-prefix matching. IAPRepair incurs negligible TCAM overhead.

\begin{table}[htbp]
\centering
\caption{Tofino ASIC Resource Usage}
\begin{tabular}{ccccc}
    \toprule 
     Stage & SRAM & Hash Bit & TCAM & VLIW Instr \\
    \midrule 
    0   & 5.00\%   & 7.93\%  & 0.00\% & 15.63\% \\ 
    1   & 1.25\%   & 2.10\%  & 0.00\% & 9.38\% \\ 
    2   & 6.25\%   & 13.46\% & 0.00\% & 9.38\% \\ 
    3   & 15.00\%  & 12.50\% & 0.00\% & 9.38\% \\ 
    4   & 15.00\%  & 12.50\% & 0.00\% & 9.38\% \\ 
    5   & 13.75\%  & 12.50\% & 0.00\% & 9.38\% \\ 
    6   & 15.00\%  & 12.50\% & 0.00\% & 9.38\% \\ 
    7   & 15.00\%  & 12.50\% & 0.00\% & 9.38\% \\ 
    8   & 15.00\%  & 12.50\% & 0.00\% & 9.38\% \\ 
    9   & 13.75\%  & 11.78\% & 0.00\% & 15.62\% \\ 
    10  & 1.25\%   & 2.40\%  & 0.00\% & 6.25\% \\ 
    \bottomrule 
    \end{tabular}
    \label{tab_exp09}
\end{table}

Experiment 10 (Algorithm runtime): The number of nodes and the number of STF blocks are two critical factors determining the runtime of IAPRepair. Table \ref{tab_node_runtime} and \ref{tab_stripe_runtime} respectively present the impact of these two factors on runtime, with the number of nodes varying from 20 to 100, and the number of STF blocks ranging from 200 to 1000. The time overhead of IAPRepair consistently outperforms that of SelectiveEC and FastPR. For various numbers of nodes, the runtime of IAPRepair is reduced by at least 17.71\% compared to other methods. For different numbers of STF blocks, IAPRepair achieves a minimum reduction of at least 18.18\% in runtime compared to other methods. The rationale is that SelectiveEC and FastPR require repeated invocations of the maximum flow algorithm for block selection, which incurs high time complexity. In contrast, IAPRepair simply computes block scores based on the distribution of STF stripes across nodes, leading to low computational overhead. 


\begin{table}[t]
\centering
\caption{Impact of the number of nodes on runtime}
\begin{tabular}{cccc}
    \toprule 
    \# nodes & IAPRepair & SelectiveEC & FastPR \\
    \midrule 
    20  & 0.29s  & 0.41s  & 0.99s  \\ 
    40  & 0.47s  & 0.61s  & 4.03s  \\ 
    60  & 0.62s  & 0.79s  & 8.51s  \\ 
    80  & 0.79s  & 0.96s  & 33.96s \\ 
    100 & 0.93s  & 1.16s  & 41.08s \\ 
    \bottomrule 
    \end{tabular}
    \label{tab_node_runtime}
\end{table}


\begin{table}[t]
\centering
\caption{Impact of the number of blocks on runtime}
\begin{tabular}{cccc}
    \toprule 
    \# blocks & IAPRepair & SelectiveEC & FastPR \\
    \midrule 
    200  & 0.09s  & 0.11s  & 0.27s  \\ 
    400  & 0.17s  & 0.25s  & 0.53s  \\ 
    600  & 0.31s  & 0.43s  & 0.79s  \\ 
    800  & 0.42s  & 0.56s  & 1.06s  \\ 
    1000 & 0.56s  & 0.71s  & 1.33s  \\ 
    \bottomrule 
    \end{tabular}
    \label{tab_stripe_runtime}
\end{table}

In production storage systems, a node contains more than 1.5 million blocks \cite{ren2017realistic, wang2019dayu}. These blocks can be partitioned into multiple smaller subsets and repaired sequentially using a pipelined scheduling-transmission approach. IAPRepair features a lightweight computational overhead (less than 1 second). Its runtime can be hidden within the data transfer time of the previous subset, thus having negligible impact on repair performance.

\section{Conclusion}
\label{sec:Conclusion}
The proposed IAPRepair performs proactive repair on the STF node based on accurate node failure prediction. We first analyze the factors influencing proactive repair. Subsequently, by employing a rational method for STF stripes partitioning and leveraging in-network aggregation for STF blocks reconstruction, we achieve fully bandwidth utilization and enhance the efficiency of proactive repair. Prototype system testing and large-scale simulations demonstrate that IAPRepair significantly reduces data repair time compared to existing methods.

\section*{Acknowledgements}

The authors would like to thank the associate editor and the anonymous reviewers for their valuable comments. This work was supported by the National Natural Science Foundation of China (Grant No: 61861013, 62162018), Guangxi Key Research and Development Project(Grant No: 2024AB24072) and the Project for Enhancing Young, Middle-aged Teacher's Research Basis Ability in Colleges of Guangxi (Grant No: 2023KY0814).

\bibliographystyle{IEEEtran}
\bibliography{sample-base2}
\vspace{-1em}

\end{document}